\documentclass[10pt,twocolumn]{article}

\usepackage[dvips,letterpaper,top=0.5in, bottom=0.5in, left=0.75in, right=0.5in,includefoot,heightrounded]{geometry}

\usepackage{srcltx}

\usepackage{amsmath}
\usepackage{amssymb,amsfonts}

\usepackage{abstract}

\usepackage{epsfig}
\usepackage[usenames,dvipsnames]{color}
\usepackage[usenames,dvipsnames]{xcolor}
\usepackage{subfigure}

\usepackage{booktabs}

\usepackage{setspace}
\usepackage{flushend}
\usepackage{url}

\usepackage{multirow}
\usepackage{array}

\usepackage[short,12hr]{datetime}

\usepackage{hyperref}

\usepackage[sc,tiny,center]{titlesec}
       \usepackage{mathptmx}

\newtheorem{definition}{Definition}

\newtheorem{lemma}{Lemma}

\newcommand{\GI}{\operatorname{GI}}
\newcommand{\GF}{\operatorname{GF}}

\newcommand{\cas}{\operatorname{cas}}

\def\QED{\mbox{$\square$}}
\def\proof{\noindent{\it Proof:~}}
\def\endproof{\hspace*{\fill}~\QED\par\endtrivlist\unskip}



\title{%
Fragile Watermarking Using Finite Field Trigonometrical Transforms
}

\author{%
R. J. Cintra%
\thanks{%
R. J. Cintra 
was with 
the Department of Electrical and Computer Engineering,
University of Calgary, Calgary, Alberta, Canada.
He is currently
with
the Signal Processing Group,
Departamento de Estat\'{\i}stica, 
Universidade Federal de Pernambuco.
E-mail: \protect\url{rjdsc@de.ufpe.br}
}
\and
V.~S.~Dimitrov\thanks{%
V.~S.~Dimitrov
with the Department of Electrical and Computer Engineering,
University of Calgary, Calgary, Alberta, Canada.}
\and
H. M. de Oliveira%
\thanks{%
H. M. de Oliveira
was with 
Departamento de Eletr\^onica e Sistemas,
Universidade Federal de Pernambuco (UFPE).
He is currently
with
the Signal Processing Group,
Departamento de Estat\'{\i}stica, 
Universidade Federal de Pernambuco.
}
\and
R. M. Campello de Souza%
\thanks{%
R. M. Campello de Souza
is with 
the Communications Research Group,
Departamento de Eletr\^onica e Sistemas,
Universidade Federal de Pernambuco.
}
}

\date{\today\ @ \currenttime}
\date{}

\newcommand{\myabstract}{%
Fragile digital watermarking has been applied for authentication and alteration detection
in images.
Utilizing the
cosine and Hartley transforms over finite fields,
a new transform domain fragile watermarking scheme is introduced.
A watermark is embedded into a host image via a blockwise application of
two-dimensional finite field cosine or Hartley transforms.
Additionally, the considered finite field transforms are adjusted to be number theoretic transforms,
appropriate for error-free calculation.
The employed technique can provide invisible fragile watermarking for authentication systems
with tamper location capability.
It is shown that the choice of the finite field characteristic is
pivotal to obtain perceptually invisible watermarked images.
It is also shown that the generated watermarked images
can be used as publicly available signature data for authentication purposes.
}

\newcommand{\mykeywords}{%
Fragile watermarking;
number theoretic transforms; 
finite field trigonometry
}

\begin{document}

\makeatletter
\if@twocolumn

\twocolumn[%
  \maketitle
  \begin{onecolabstract}
    \myabstract
  \end{onecolabstract}
  \begin{center}
    \small
    \textbf{Keywords}
    \medskip
    \linebreak
    \mykeywords
  \end{center}
  \bigskip
]
\saythanks

\else

  \maketitle
  \begin{abstract}
    \myabstract
  \end{abstract}
  \begin{center}
    \small
    \textbf{Keywords}
    \medskip
    \linebreak
    \mykeywords
  \end{center}
  \bigskip
  \onehalfspacing
\fi

\section{Introduction}

Finite field transforms
link two sequences of data
according to a general pair of relationships
given by
\begin{align}
V_k
&=
\sum_{i=0}^{N-1}
v_i
K(i,k,\zeta),
\quad k = 0,1,\ldots,N-1, \\
v_i
&=
\sum_{k=0}^{N-1}
V_k
K^{-1}(i,k,\zeta),
\quad i = 0,1,\ldots,N-1,
\end{align}
where
$\mathbf{v}=(v_0\ v_1 \cdots v_{N-1})$ and
$\mathbf{V}=(V_0\ V_1 \cdots V_{N-1})$ are data vectors of length~$N$
with elements defined over a certain
Galois field;
$K(\cdot,\cdot,\cdot)$ and
$K^{-1}(\cdot,\cdot,\cdot)$
are the forward and inverse transformation kernels, respectively;
and
$\zeta$ is a fixed element of the given Galois field~\cite{thomas1983independent}.

In this framework, the concept of number theoretic transforms arises.
If, for each vector $\mathbf{v}$ with components defined over $\GF(p)$,
its associated transform pair $\mathbf{V}$ has also components defined over $\GF(p)$,
then the transformation is said to be a number theoretic transform~(NTT)~\cite{campello2000numericas}.
Number theoretic transforms have been subject of much interest,
because of their capability of exact calculation.
Being all operations required by an NTT performed
in a finite field,
there are no errors due to rounding or truncation.
Consequently,
in principle,
the possibility of error-propagation is eliminated.

The first proposed NTT, due to Pollard~\cite{pollard1971fast},
is a Fourier-like transform
often called finite field Fourier transform (FFFT).
Since its introduction,
the FFFT
was submitted to several
generalizations~\cite{dimitrov1994fermat,dimitrov1992generalized}
and
further methods
were devised~\cite{dimitrov1995golden,hong1993hartley,boussakta1992new,%
souza2004ffdct,souza1999ffht,campello2000numericas,campello1998trigonometry,campello2000hntt,silva2000transformadas}.

Standard number theoretic transforms
have been employed in
different frameworks,
such as:
(i)~fast calculation of convolutions and correlations~\cite{reed1975convolution,agarwal1975convolution,martens1984polynomial};
(ii)~algebraic coding theory~\cite{mcwilliams1988error};
and
(iii)~very large number multiplication~\cite{schonhage1971multiplikation}.
In the past years,
it was observed a broadening
of the NTT range of applications.
Other topics of research have also been benefited from the
utilization of NTT approaches.
Just to illustrate some of them,
one could cite:
(i)~solving Toeplitz system of equations~\cite{hsue1995toeplitz};
(ii)~speech coding problems~\cite{madre2003pitch};
and
(iii)~fast matrix multiplication~\cite{yagle1995matrix}.
Two-dimensional number theoretic transforms were subject to a comprehensive exposition in~\cite{julliendimitrov1996two}.
Additionally,
image processing problems,
involving motion estimation~\cite{toivonen2002motion}
and geometrical rotation~\cite{kriz1980rotation},
were also addressed.
In~\cite{tamori2002fragile}, Tamori~\emph{et al.}
suggested the use of the Fourier-based NTT
to provide a fragile watermarking scheme.
In the current work, this specific technique is denominated
Tamori-Aoki-Yamamoto scheme.

Generally, a watermarking operation consists of encapsulating a given information
(e.g., an image)
into raw image data.
The former is the watermark, the latter is the image to be watermarked.
Frequently, the watermark is required to be perceptually transparent.
Invisible watermarks are typically categorized in two different models:
robust or fragile watermarking.
Depending on the purpose of the watermarking,
robust or fragile schemes are chosen.
Robust watermarks are designed to endure
a variety of data manipulations,
such as adjustments of the compression ratio,
filtering, cropping, or scaling.
Therefore, robust watermarks are
suitable in situations involving
misappropriation of data,
such as
ownership assertion
and copyright enforcement~\cite{lin1999review}.
Despite of this, because of its robustness,
this class of watermarking methodology
is in part
ineffectual for recognizing tampering.

On the other hand,
fragile watermarking schemes
can furnish the necessary tools
for authentication
and integrity
attestation.
In fact, fragile watermarking techniques are supposed to detect tampering
and
determine the identity of the data originator
with high probability~\cite{lin1999review,george1999watermarking}.
However, fragile watermarks are not adequate in situations that require
copyright verification.
Since minimal image alterations are expected to promptly damage fragile watermarks,
fragile watermarking can not properly deal with misappropriation issues.

Several fragile watermarking techniques have been proposed,
usually, with the purpose of still image authentication.
A glimpse of the current investigations
embraces
works on
simultaneously robust and fragile systems~\cite{lu2001multipurpose};
authentication methods for {JPEG} images~\cite{li2004digital,lu2003fragile};
protection of video communications~\cite{chen2005video};
and
tamper detection with the aid of wavelet decompositions~\cite{hu2002multiresolution,kundur1999telltale}.
Some non-conventional applications have also
been reported, such as
blind estimation of the quality of communication links~\cite{campisi2003blind},
and
image quality assessment~\cite{zheng2003image}.

In account of the above,
the number theoretic transforms arise as a natural tool to provide fragile
watermarking methods.
Since 
(i) the NTT domain has no physical meaning, 
such as harmonic content,
and
(ii) the concept of energy over a finite field is not clear,
any perturbation on a data sequence
produces a dramatic alteration of
its associated number theoretic transformed sequence.
Moreover,
all computations required by an NTT
are integer modular arithmetic,
which can be efficiently implemented.

The aim of this paper is twofold.
First, an amplification of the Tamori-Aoki-Yamamoto fragile watermarking scheme is sought~\cite{tamori2002fragile}.
To explore this line, finite field trigonometry and trigonometrical transforms
are employed~\cite{souza2004ffdct,souza1999ffht,campello2000numericas,campello1998trigonometry,campello2000hntt,silva2000transformadas}.
Second,
a new mode of operation for the
discussed
watermarking scheme is suggested.
The original method by Tamori~\emph{et al.} is categorized as a private watermarking system.
In the present study, the proposal of a signature method with the ability of tamper detection and location
is made.

The rest of the paper is organized as follows.
Section~\ref{sectionfftt} is devoted
to the finite field trigonometry theory.
Central aspects of finite field transforms are also outlined.
Mainly, the focus is directed to the finite field cosine transform (FFCT)
and to the finite field Hartley transform (FFHT).
In Section~\ref{sectiontayschem},
the fragile watermarking technique proposed in~\cite{tamori2002fragile}
is described.
Then,
a new \emph{modus operandi} for the discussed watermarking scheme
is elaborated in Section~\ref{sectionsignature}.
Finally,
computational
results are presented in
Section~\ref{sectioncomputational}.
Concluding remarks are given in Section~\ref{sectionconclusion}.

\section{Finite Field Trigonometrical Transforms}
\label{sectionfftt}

In a series of papers
\cite{souza1999ffht,souza2004ffdct,campello2000hntt,campello1998trigonometry,silva2000transformadas}
by Campello \mbox{de Souza} and collaborators,
a trigonometry over finite fields was derived.
Equipped with such trigonometrical tools, it is possible to
define finite field transforms other than
Pollard's Fourier transform over finite fields~\cite{pollard1971fast}.
In particular,
the finite field trigonometry successfully offers
a formalism to encompass
the finite field cosine transform  and
the finite field Hartley transform.

In this section, the theory of finite field trigonometry is briefly discussed and
its major properties are emphasized.
This is necessary to pave
the way for subsequent definitions of the
finite field trigonometrical transforms.

Let $\GF(p)$ be a Galois field
with odd characteristic~$p$.
If $p \equiv 3 \pmod{4}$,
then it is possible to construct an extension field $\GF(p^2)$
using the solution of the irreducible polynomial
$x^2+1=0$~\cite{dimitrov1995golden}.
In addition,
this extension field is isomorphic to the Gaussian integer field $\GI(p)
= \{ a + jb :  a,b\in\GF(p) \}$, where
$j^2 \equiv -1 \pmod{p}$.
Furthermore,
compared to the usual complex numbers,
the Gaussian integers enjoy similar arithmetic operation rules.

For a given element $\zeta\in\GI(p)$ with multiplicative order $N$,
the trigonometrical cosine and sine functions are defined
by~\cite{souza1999ffht,campello1998trigonometry}
\begin{equation}
\label{eq.cosine.definition}
\cos(i)
\triangleq
\frac{\zeta^i + \zeta^{-i}}{2},
\quad
i = 0, 1, \ldots, N-1,
\end{equation}
\begin{equation}
\sin(i)
\triangleq
\frac{\zeta^i - \zeta^{-i}}{2j},
\quad
i = 0, 1, \ldots, N-1,
\end{equation}
respectively.
Remarkably,
for each $\zeta$ in $\GI(p)$,
the trigonometrical functions define a
different mapping.
Moreover,
unlike their real field counterparts,
the finite field cosine and sine
can assume complex values,
since $\zeta$ is a Gaussian integer.

The following definition and lemma provide
means to circumvent the need of complex arithmetic.

\begin{definition}[Unimodularity~\cite{silva2000transformadas}]
An element $a+jb\in \GI(p)$ is unimodular if
$a^2 + b^2  \equiv 1 \pmod{p}$.
\endproof
\end{definition}
In other words,
unimodular elements have unitary Gaussian integer norm,
denoted by $n(a+jb)\triangleq a^2 + b^2 \pmod{p}$~\cite{hardy1945introduction}.

Besides, if $\zeta$ has unitary norm,
then the computation of trigonometrical functions
are simplified,
as it is shown in the following lemma.
Let the real and the imaginary parts of a Gaussian integer $a+jb$,
where $a,b\in\GF(p)$,
be denoted by $\Re(a+jb)=a$ and $\Im(a+jb)=b$, respectively.

\begin{lemma}[Real Cosine and Sine~\cite{campello2000numericas}]
If
$\zeta\in \GI(p)$ is unimodular with multiplicative order~$N$,
then
$\cos(i) = \Re(\zeta^i)$ and
$\sin(i) = \Im(\zeta^i)$,
$i = 0, 1, \ldots, N-1$.
\end{lemma}
\proof
Manipulating the cosine function expression (Equation~\ref{eq.cosine.definition}) yields
\begin{align}
\cos(i) & = \frac{\zeta^i + \zeta^{-i}}{2} \\
&=\frac{\zeta^i + (\zeta^i)^{-1}}{2}.
\end{align}
Let $\zeta^i = a + j b$, where $a,b\in\GF(p)$.
Then, by usual arithmetical operations, the inverse of $\zeta^i$ is given by
\begin{align}
(\zeta^i)^{-1}
=
\frac{a - jb}{a^2 + b^2}
=
\frac{a - jb}{n(\zeta^i)}.
\end{align}
On account of the fact that
$n(\cdot)$ has a multiplicative property~\cite{hardy1945introduction},
$n(\zeta^i) = [n(\zeta)]^i = 1$.
Thus, the inverse of $\zeta^i$ is equal to its complex conjugate
$a-jb$.
Therefore, it yields
\begin{align}
\cos(i) & = \frac{(a+jb) + (a-jb)}{2} \\
&= a = \Re(\zeta^i).
\end{align}
A comparable derivation can be obtained for the sine function.
\endproof

This result is of paramount importance.
Once a unimodular element $\zeta$ is chosen,
any arithmetic manipulation involving finite field cosine or sine
functions results in quantities defined over the ground field $\GF(p)$.
Therefore,
the unimodularity of $\zeta$ is a sufficient
condition
for the definition of number theoretic transforms that utilize the finite field
cosine or sine.
Under this condition,
two consequences are accomplished:
(i) the finite field trigonometrical functions induce no complex arithmetic operations
and
(ii) all related computations are performed by means of modular integer arithmetic.
In terms of computational implementation, 
these two properties are significant.

\subsection{Finite Field Cosine Transform}

Early investigations on the finite field discrete cosine transform were
reported in~\cite{julliendimitrov1996two}.
Independently,
an extended study of the FFCT,
based on  the finite field trigonometry,
was introduced in~\cite{souza2004ffdct}.
A comprehensive derivation
and detailed proofs of related FFCT theorems are described in~\cite{souza2004ffdct}.

Generally, the FFCT is a transformation that
provides a spectrum that is not in $\GF(p)$,
thereby exhibiting a complex nature.
Nevertheless, for carefully selected unimodular $\zeta$ values,
it is possible to ensure that the FFCT spectrum
assumes purely real quantities.
Consequently,
the FFCT becomes an NTT.
In the present study,
only the NTT case is of interest.
The following simplified presentation of the FFCT is formulated to guarantee
a real spectrum.

\begin{definition}[FFCT~\cite{souza2004ffdct,julliendimitrov1996two}]
For a given unimodular element $\zeta\in\GI(p)$ with multiplicative order $4N$,
the
FFCT
maps an
$N$-dimensional vector
with elements $v_i \in\GF(p)$, $i = 0, 1, \ldots, N-1$,
into
another vector with components $V_k \in\GF(p)$, $k = 0, 1, \ldots, N-1$
according to
\begin{equation}
\label{ffct}
V_k \triangleq
\sum_{i=0}^{N-1}
2 v_i
\cos
\big(
(2i+1)k
\big),
\end{equation}
for $k = 0, 1, \ldots, N-1$.
\endproof
\end{definition}
An inversion formula can be derived and
is given by the following expression~\cite{souza2004ffdct,julliendimitrov1996two}:
\begin{equation}
\label{iffct}
v_i =
\frac{1}{N \pmod{p}}
\sum_{k=0}^{N-1}
a_k
V_k
\cos\big( (2i+1)k \big),
\end{equation}
for $i = 0, 1, \ldots, N-1$.
The auxiliary sequence $\mathbf{a}$ is defined by
\begin{equation}
\label{auxiliar}
a_i
=
\begin{cases}
2^{-1} \pmod{p}, & \text{if $i=0$}, \\
1, & \text{otherwise},
\end{cases}
\end{equation}
for $i = 0, 1, \ldots, N-1$.
Table~\ref{tab:zetas:ffct} brings a list of adequate unimodular $\zeta$ elements to be used
in the definition of cosine and sine functions
in order
to make the FFCT act as an NTT.

\begin{table}
\caption{Suitable values of $\zeta$ and the associated transform blocklength $N$ 
for the FFCT over several finite fields $\GF(p)$}
\label{tab:zetas:ffct}
\begin{minipage}{\linewidth}
\centering
{\scriptsize
\begin{tabular}{ccm{5cm}}
\hline
$p$ & $N$ &  $\zeta$ \\
\hline
\multirow{1}{0.5cm}{7}
&    2 & $\pm 2 \pm 2j$ \\
\hline
\multirow{1}{0.5cm}{11}
&    3 & $\pm 3 \pm 5j$ \\
\hline
\multirow{1}{0.5cm}{19}
&    5 & $\pm 4 \pm 2j$, $\pm 3 \pm 7j$   \\
\hline
\multirow{3}{0.5cm}{23}
&    2 & $\pm 9 \pm 9j$ \\
&    3 & $\pm 8 \pm 11j$ \\
&    6 & $\pm 10 \pm 4j$, $\pm 4 \pm 10j$ \\
\hline
\multirow{3}{0.5cm}{31}
&    2 & $\pm 4 \pm4j$ \\
&    4 & $\pm 13\pm 7j$, $\pm7 \pm 13j$ \\
&    8 & $\pm 11 \pm 2j$, $\pm 10 \pm 5j$, $\pm 5 \pm 10j$, $\pm 2 \pm 11j$  \\
\hline
\multirow{1}{0.5cm}{43}
&    11 & $\pm 9 \pm 7j$, $\pm 3 \pm 11j$, $\pm 13 \pm 2j$, $\pm 8 \pm 18j$, $\pm 20 \pm 17j$ \\
\hline
\multirow{5}{0.5cm}{47}
&    2 & $\pm 20 \pm 20j$ \\
&    3 & $\pm 6 \pm 23j$ \\
&    4 & $\pm 22 \pm 9j$, $\pm 9 \pm 22j$ \\
&    6 & $\pm 16 \pm 11j$, $\pm 11 \pm 16j$ \\
&    12 & $\pm 19 \pm 4j$, $\pm 4 \pm 19j$, $\pm 18 \pm 10j$, $\pm 10 \pm 18j$  \\
\hline
\multirow{7}{0.5cm}{127\footnote{Selected $\zeta$ values only.}}
&    2 & $\pm 8\pm 8j$ \\
&    4 & $\pm 24 \pm 21j$, $\pm 21 \pm 24j$ \\
&    8 & $\pm 30 \pm 25j$, $\pm 59 \pm 40j$  \smallskip \\ 
&    16 & $\pm 29 \pm 7j$, $\pm 41 \pm 15j$,  $\pm 49 \pm 34j$,  $\pm 60 \pm 46j$, $\pm 67 \pm 46j$ \smallskip \\
&    32 & $\pm 22 \pm 5j$, $\pm 23 \pm 19j$,  $\pm 39 \pm 2j$,   $\pm 2 \pm 39j$,  $\pm 38 \pm 9j$,  $\pm 45 \pm 32j$ \\
\hline
\end{tabular}
}
\end{minipage}
\end{table}

\subsection{Two-dimensional FFCT}

As the kernel of FFCT is separable,
it follows that
the two-dimensional
FFCT can be performed by successive calls of the one-dimensional FFCT
applied to the rows of the image data; then to the columns of the resulting intermediate
calculation.
Invoking the Equation~\ref{ffct},
the two-dimensional finite field cosine transform can be synthesized in matrix form.
Let~$\mathbf{C}$ be the FFCT matrix,
whose elements are given by
\begin{equation}
c_{i,k} =  2\cos\big((2i+1)k\big),
\quad i,k = 0, 1, \ldots, N-1.
\end{equation}
The two-dimensional FFCT of an $N\times N$ image data $\mathbf{D}$  can be simply written as
\begin{equation}
\hat{\mathbf{D}} = \mathbf{C}\cdot\mathbf{D}\cdot\mathbf{C}^T,
\end{equation}
where the superscript~$T$ is the transposition operation.
The matrix formulation for the inverse transformation
can be obtained in a similar way:
\begin{equation}
\mathbf{D} = \mathbf{C}^{-1}\cdot\hat{\mathbf{D}}\cdot(\mathbf{C}^{-1})^T,
\end{equation}
where,
according to Equation~\ref{iffct},
the elements of $\mathbf{C}^{-1}$ are given by
\begin{equation}
c'_{i,k} =
\frac{1}{N \pmod{p}}
a_k
\cos\big( (2i+1)k \big),
\quad i,k = 0, 1, \ldots, N-1,
\end{equation}
and the quantities $a_k$ are defined in Equation~\ref{auxiliar}.

\subsection{Finite Field Hartley Transform}

Introduced independently in~\cite{hong1993hartley,campello1998trigonometry},
the definition of the FFHT resembles the formalism of
the conventional discrete Hartley transform (DHT)~\cite{bracewell1986hartleytransform}.

\begin{definition}[FFHT~\cite{campello1998trigonometry}]
For a given $\zeta\in\GI(p)$ with multiplicative order $N$,
the FFHT relates two $N$-dimensional vectors $\mathbf{v}$ and $\mathbf{V}$,
according to the following expression
\begin{equation}
V_k \triangleq \sum_{i=0}^{N-1} v_i \cas(ik), \quad k = 0, 1, \ldots, N-1,
\end{equation}
where
\begin{align}
\cas(i) &\triangleq \cos(i) + \sin(i) \\
        &=\frac{(1-j)\zeta^i + (1+j)\zeta^{-i}}{2}
\end{align}
is the finite field version of the Hartley function~\cite{bracewell1986hartleytransform}.
\endproof
\end{definition}
Consonant with the DHT theory, 
apart from the scaling factor $N^{-1}\pmod{p}$,
the FFHT is a symmetrical
transformation
and its inversion formula is given by~\cite{campello2000hntt,campello1998trigonometry,campello2000numericas,souza1999ffht}
\begin{equation}
v_i =
\frac{1}{N\pmod{p}}
\sum_{k=0}^{N-1}
V_k
\cas(ik),
\quad
i = 0, 1, \ldots, N-1.
\end{equation}

Similar to the FFCT, the FFHT can exhibit a complex spectrum.
Nevertheless,
a judicious choice of $\zeta$ can effectively prevent this behavior,
making the components of the finite field Hartley spectrum to be real.
Again, unimodular elements are taken into consideration.
Table~\ref{tab:zetas:ffht} lists
some unimodular $\zeta$ elements for several fields,
and the associated FFHT blocklength.

The FFHT can be written in terms of matrices.
Let~$\mathbf{H}$ be the finite field Hartley matrix,
whose elements $h_{i,k}$ are
given by
\begin{equation}
h_{i,k}= \cas(ik),
\quad i,k = 0, 1, \ldots, N-1.
\end{equation}
Consequently, the forward and inverse FFHT are expressed by
\begin{align}
\mathbf{V} &= \mathbf{H}\cdot \mathbf{v}, \\
\mathbf{v} &= \frac{1}{N\pmod{p}}\mathbf{H}\cdot \mathbf{V},
\end{align}
respectively.

\begin{table}
\caption{Suitable values of $\zeta$ and the associated transform blocklength $N$ 
for the FFHT over several finite fields $\GF(p)$}
\label{tab:zetas:ffht}
\begin{minipage}{\linewidth}
\centering
{\scriptsize
\begin{tabular}{ccm{5cm}}
\hline
$p$ & $N$ & $\zeta$  \\
\hline
\multirow{2}{0.5cm}{3}
&    2  &       $2$ \\
&    4  &    $j$, $2j$    \\
\hline
\multirow{3}{0.5cm}{7}
&       2 &    $6$ \\
&       4 &    $\pm j$ \\
&       8 &    $\pm2\pm2j$ \\
\hline
\multirow{5}{0.5cm}{11}
&      2 &    $10$ \\
&       3 &    $5\pm3j$ \\
&       4 &    $\pm j$ \\
&       6 &    $6\pm 3j$ \\
&       12 &    $\pm3\pm 5j$ \\
\hline
\multirow{5}{0.5cm}{19}
& 2 & $18$ \\
& 4 & $\pm j$ \\
& 5 & $2\pm4j$, $7\pm3j$ \\
& 10 & $12\pm3j$, $17\pm4j$ \\
& 20 & $\pm3\pm7j$, $\pm4\pm2j$ \\
\hline
\multirow{7}{0.5cm}{23}
& 2 & $22$ \\
& 3 & $11\pm 8j$ \\
& 4 & $\pm j$ \\
& 6 & $12\pm 8j$ \\
& 8 & $\pm9\pm 9j$ \\
&      12 &    $\pm8\pm 11j$ \\
&      24 &    $\pm4\pm 10j$, $\pm10\pm 4j$ \\
\hline
\multirow{5}{0.5cm}{31}
& 2 & $30$ \\
& 4 & $\pm j$ \\
& 8 & $\pm4\pm 4j$ \\
& 16 & $\pm7\pm 13j$, $\pm13\pm 7j$ \\
& 32 & $\pm2\pm 11j$, $\pm5\pm 10j$, $\pm10\pm 5j$, $\pm11\pm 2j$ \\
\hline
\multirow{5}{0.5cm}{43}
& 2 & $42$ \\
& 4 &    $\pm j$ \\
& 11 & $2\pm 13j$, $7\pm 9j$, $11\pm 3j$, $18\pm 8j$, $26\pm 20j$ \\
& 22 & $17\pm 20j$, $25\pm 8j$, $32\pm 3j$, $36\pm 9j$, $41\pm 13j$ \\ 
& 44 & $3\pm 11j$, $\pm 8\pm 18j$, $\pm 9\pm 7j$, $\pm 13\pm 2j$, $\pm 20\pm 17j$, $40\pm 11j$    \\
\hline
\multirow{9}{0.5cm}{47}
&    2 & $46$ \\
&    3 & $23\pm 6j$ \\
&    4 & $\pm j$ \\
&    6 & $24\pm 6j$ \\
&    8 & $20\pm 20j$ \\ 
&    12 & $6\pm 23j$ \\ 
&    16 & $\pm 22\pm 9j$, $\pm 9\pm 22j$ \\  
&    24 & $\pm 16\pm 11j$ \\ 
&    48 & $\pm 19\pm 4j$, $\pm 4\pm 19j$, $\pm 18\pm 10j$, $\pm 10\pm 18j$ \\
\hline
\multirow{9}{0.5cm}{127\footnote{Selected $\zeta$ values only.}}
&    2 & $126$ \\
&    4 & $\pm j$ \\
&    8 & $\pm 8\pm 8j$ \\
&   16 & $\pm 24 \pm 21j$, $\pm 21 \pm 24j$ \\
&   32 & $\pm 30 \pm 25j$, $\pm 59 \pm 40j$  \smallskip \\ 
&   64 & $\pm 29 \pm 7j$, $\pm 41 \pm 15j$,  $\pm 49 \pm 34j$,  $\pm 60 \pm 46j$, $\pm 67 \pm 46j$ \smallskip \\
&  128 & $\pm 22 \pm 5j$, $\pm 23 \pm 19j$,  $\pm 39 \pm 2j$,   $\pm 38 \pm 9j$,  $\pm 45 \pm 32j$ \\
\hline
\end{tabular}
}
\end{minipage}
\end{table}

\subsection{Two-dimensional FFHT}

Since
the FFHT lacks a separable kernel,
the two-dimensional FFHT cannot be performed by successive calls of the one-dimensional FFHT
applied to the rows of an image data; and then to its
columns~\cite{bracewell1986hartleytransform}.
In fact, the procedure for the calculation of the 2-D FFHT is analogous to the one utilized for
the computation of the 2-D DHT~\cite{bracewell1986hartleytransform}.
Initially, a temporary matrix $\mathbf{T}$ is computed,
which is given by
\begin{equation}
\mathbf{T} = \mathbf{H} \cdot \mathbf{D} \cdot \mathbf{H},
\end{equation}
where $\mathbf{D}$ is an $N\times N$ image and $\mathbf{H}$ is
the finite field Hartley transform matrix.

Subsequently, the two-dimensional FFHT of $\mathbf{D}$ is expressed by
\begin{equation}
\hat{\mathbf{D}} =
\frac{1}{2}\left( \mathbf{T} + \mathbf{T}^{\text{(c)}} + \mathbf{T}^{\text{(r)}} - \mathbf{T}^{\text{(c,r)}} \right),
\end{equation}
where
$\mathbf{T}^{\text{(c)}}$, $\mathbf{T}^{\text{(r)}}$, and $\mathbf{T}^{\text{(c,r)}}$
are built from the temporary matrix~$\mathbf{T}$.
Their elements are respectively given by
$t_{i,N-j\pmod{N}}$,
$t_{N-i\pmod{N},j}$, and
$t_{N-i\pmod{N},N-j\pmod{N}}$,
where $t_{i,j}$ are the elements of~$\mathbf{T}$,
for $i,j=0,\ldots,N-1$.

\subsection{Revisiting the FFFT}

In the light of the finite field trigonometry,
the classic finite field Fourier transform~\cite{pollard1971fast}
can be re-examined.
Mimicking the definition of the standard discrete Fourier transform,
one may consider a Fourier-like finite field transform
equipped with a kernel given by
$\cos(i) + j \sin(i)$, $i\in\GF(p)$.
This kernel suggests the following expression
\begin{align}
V_k &\triangleq
\sum_{i=0}^{N-1} v_i \big( \cos(ik) + j \sin(ik) \big) \\
&=
\sum_{i=0}^{N-1} v_i \left( \frac{\zeta^{ik}+\zeta^{-ik}}{2} + j \frac{\zeta^{ik}-\zeta^{-ik}}{2j} \right) \\
&=
\sum_{i=0}^{N-1} v_i \zeta^{ik},
\quad k = 0, 1, \ldots, N-1.
\end{align}
Because $\zeta$ is an element of $\GI(p)\cong\GF(p^2)$,
the above derivation is a simplified construction of Pollard's FFFT~\cite{pollard1971fast}.
The original FFFT definition considers
a more general field, $\GF(q^m)$,
where $q=p^r$ and $m,r$ are positive integers~\cite{blahut1979transform}.
Observe that the Fourier kernel degenerates
the
finite field trigonometrical functions into
the powers of $\zeta$.
Therefore, if $\zeta\in\GF(p)$, then the FFFT can always act as an NTT.
Otherwise, if $\zeta\in\GI(p)$, then a possibly complex spectrum can be obtained~\cite{reed1975convolution}.
The inversion formula is given by
\begin{equation}
v_i =
\frac{1}{N\pmod{p}}
\sum_{i=0}^{N-1} V_k \zeta^{-ik},
\quad i = 0, 1, \ldots, N-1.
\end{equation}
A proof of the inversion formula can be found in~\cite{blahut1979transform}.

\section{Fragile Watermarking over Finite Fields}
\label{sectiontayschem}

In this section,
a generalization of the
Tamori-Aoki-Yamamoto methodology
is presented.
In this derivation,
instead of using the Fourier-based NTT,
as suggested in~\cite{tamori2002fragile},
a general number theoretic transform is employed.

\subsection{Watermark Embedding}

Let $\mathbf{D}$ be an image data to be watermarked.
By means of modular arithmetic with respect to $p$,
the image data can
furnish its residue part~$\mathbf{D}_R$.
Therefore,
$\mathbf{D}$ can be written as
\begin{equation}
\mathbf{D} = \mathbf{D}_R + \mathbf{D}_M,
\end{equation}
where
$\mathbf{D}_R \equiv \mathbf{D} \pmod{p}$
and
$\mathbf{D}_M$ is an image containing elements that are multiples of~$p$.

The method consists of the insertion of a watermark image~$\mathbf{W}$
into the NTT spectrum of the residue of~$\mathbf{D}$.
Let $\hat{\mathbf{D}}_R$ be the \mbox{2-D}~NTT of~$\mathbf{D}_R$.
Consequently, it follows that
\begin{equation}
\hat{\mathbf{D}}_R' = \hat{\mathbf{D}}_R + \mathbf{W} \pmod{p},
\end{equation}
where $\hat{\mathbf{D}}_R'$ is the watermarked spectral contents of~$\mathbf{D}_R$.
Performing the inverse \mbox{2-D}~NTT
on $\hat{\mathbf{D}}_R'$
results in a watermarked spatial domain image~$\mathbf{D}_R'$
associated to the residue of the host image.

In view of that,
the resulting final watermarked image $\mathbf{D}'$ can be derived as follows
\begin{align}
\label{eq-watermarked}
\mathbf{D}' = \mathbf{D}_R' + \mathbf{D}_M.
\end{align}

\subsection{Watermark Extraction}

The inverse operation, watermarking extraction, can
be implemented in a similar way.
First, one may compute the modular residue
of the original and the watermarked images:
\begin{align}
\mathbf{D}_R &= \mathbf{D} \pmod{p},\\
\mathbf{D}_R' &= \mathbf{D}' \pmod{p}.
\end{align}
Both resulting residue images are submitted to
an NTT application.
Afterwards, the watermark can be recovered
from the
difference of the obtained NTT spectra:
\begin{align}
\mathbf{W} = \hat{\mathbf{D}}_R' - \hat{\mathbf{D}}_R \pmod{p},
\end{align}
where
$\hat{\mathbf{D}}_R'$ and $\hat{\mathbf{D}}_R$
are the number theoretic transform of
$\mathbf{D}_R'$ and $\mathbf{D}_R$, respectively.

\subsection{Private Watermarking Operation}

The above described
watermarking scheme
is intended to be
utilized as
a private watermarking technique for tampering detection and location,
in which the receiver must have the original image~\cite{tamori2002fragile}.
This is often the case in which the authority who marks the
data is also the interested party in verifying its integrity~\cite{lin1999review}.
Therefore, an original image $\mathbf{D}$ is watermarked to provide another image $\mathbf{D}'$,
which is transmitted.
The genuineness of $\mathbf{D}'$
is confirmed
when extracting an unaltered watermark.

\begin{figure}
\centering
\subfigure[$512\times 512$]{\includegraphics[width=0.4\linewidth,height=0.4\linewidth]{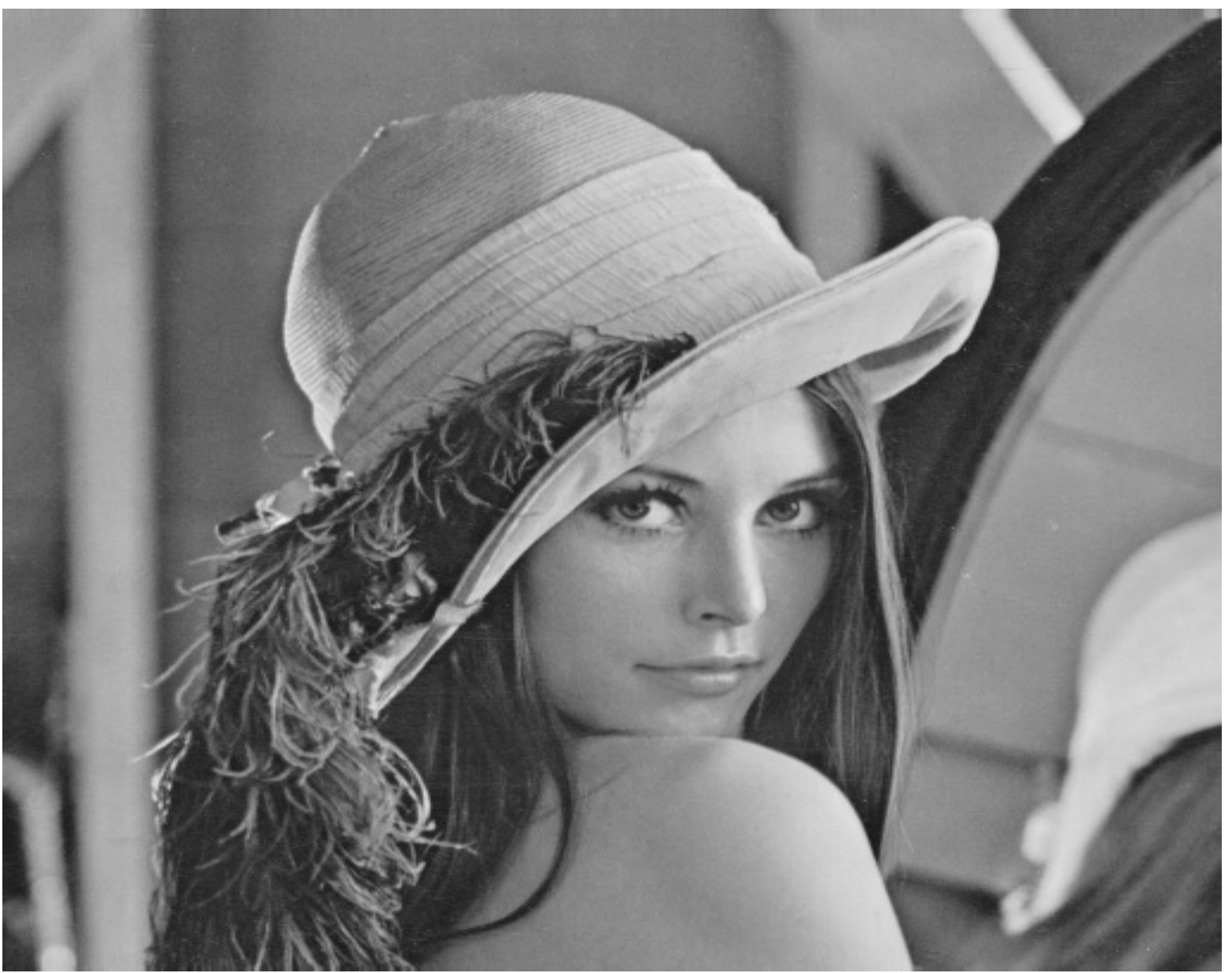}}
\subfigure[$64\times64$]{
\hspace{0.1\linewidth}
\includegraphics[width=0.2\linewidth,height=0.2\linewidth]{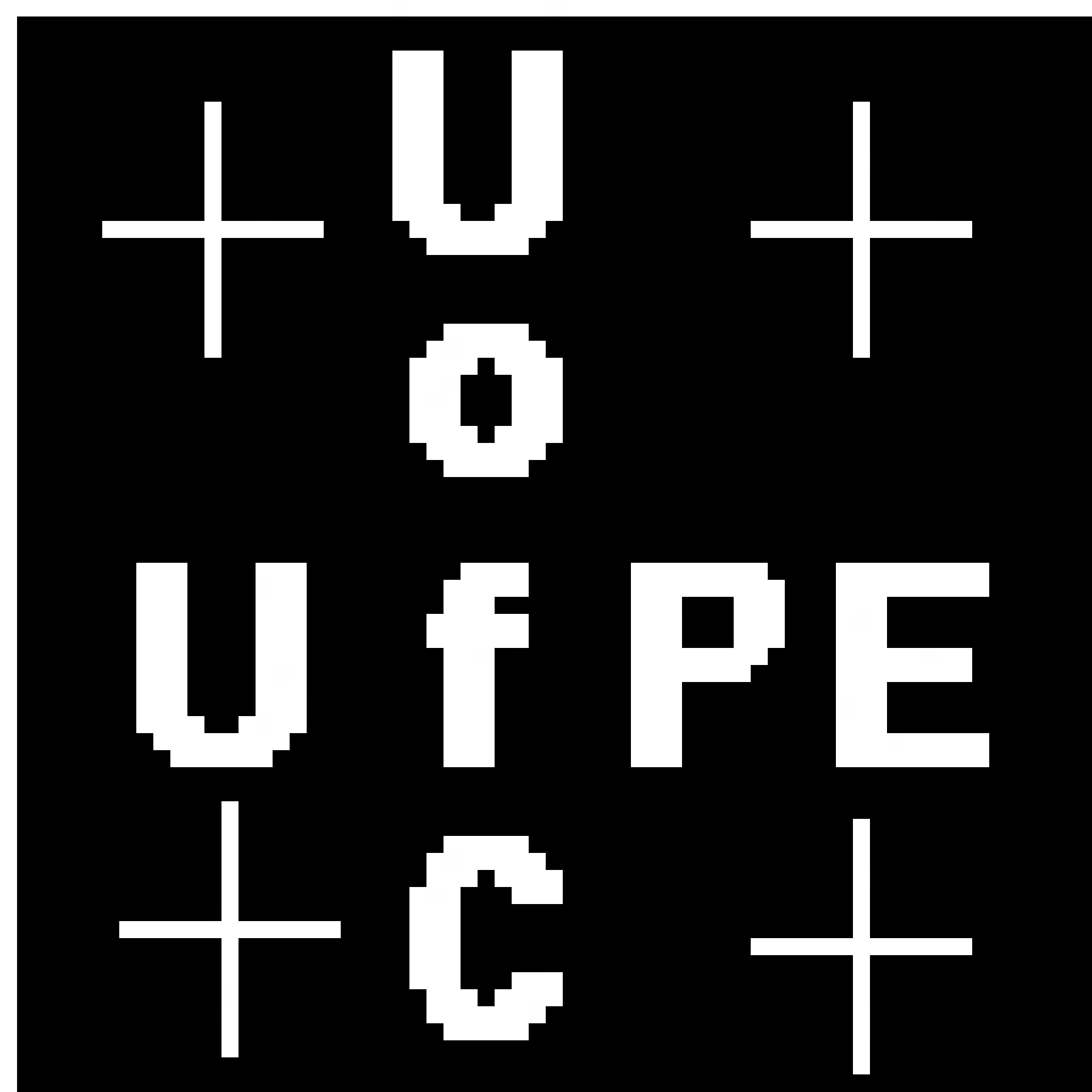} 
\hspace{0.1\linewidth}\strut
}
\\
\subfigure[$512\times 512$]{\includegraphics[width=0.4\linewidth,height=0.4\linewidth]{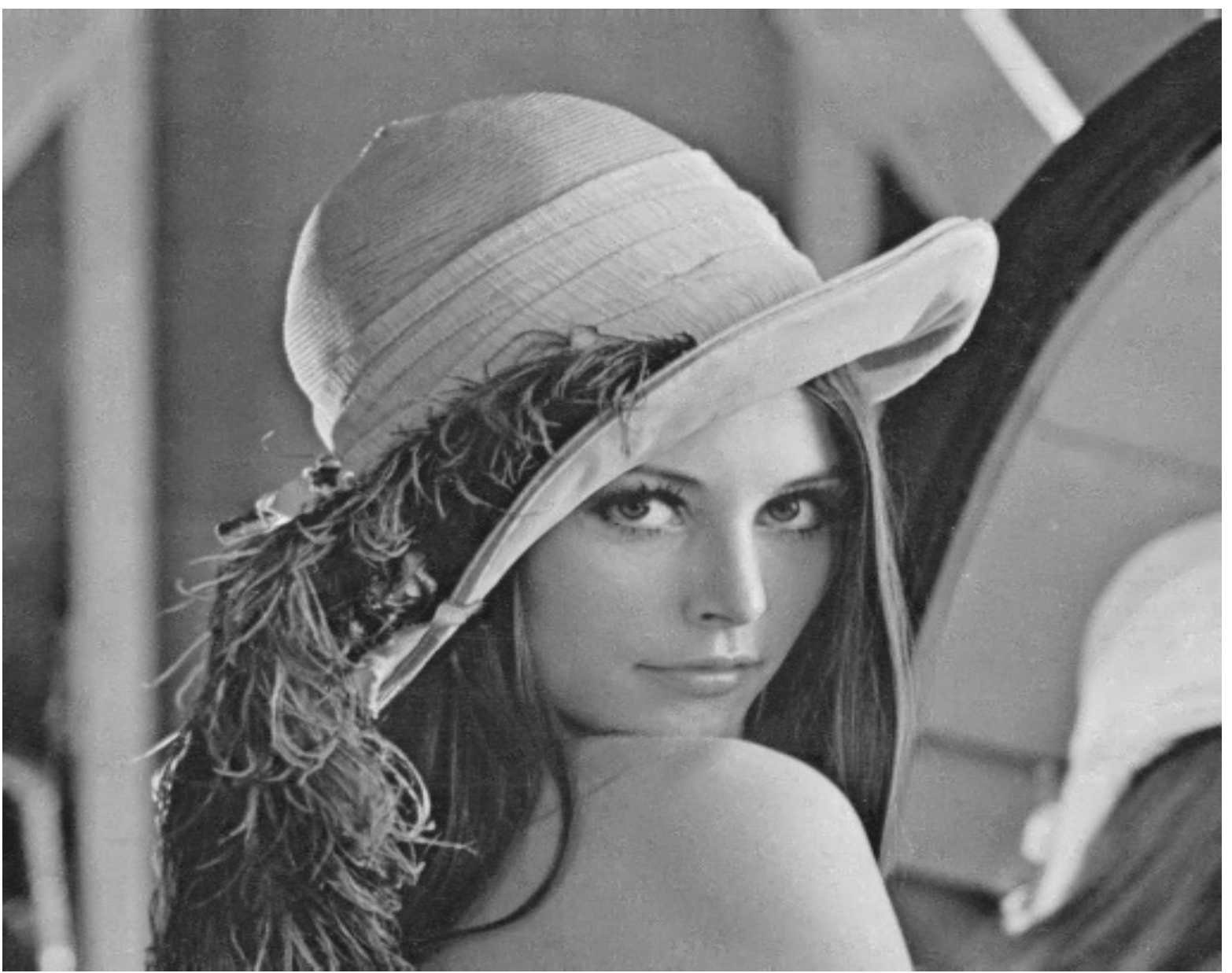} }
\subfigure[$512\times 512$]{\includegraphics[width=0.4\linewidth,height=0.4\linewidth]{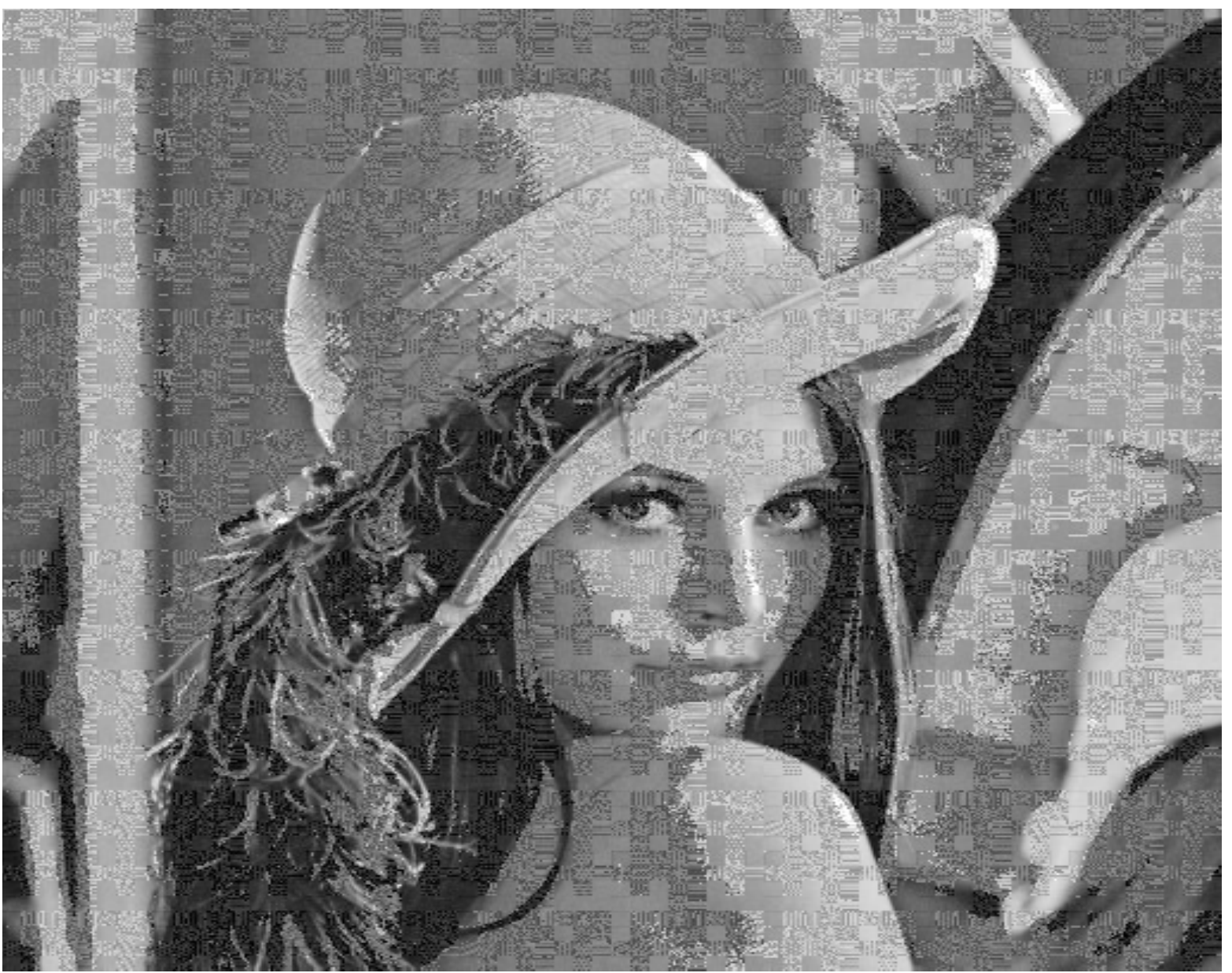} }
\caption{Private Watermarking using the FFFT.
(a)~Original image,
(b)~watermark,
(c)~watermarked image in $\GF(13)$,
(d)~watermarked image in $\GF(73)$.}
\label{fig:private:fourier}
\end{figure}

Equation~\ref{eq-watermarked}
reveals that, to produce watermarked images that are perceptually
transparent,
the dynamic range of the elements of $\mathbf{D}_R'$
must not be excessively large.
Otherwise,
watermarked images can present significant distortions.
To illustrate this behavior,
the originally proposed scheme
was employed to embed a given watermark (Figure~\ref{fig:private:fourier}(b)) into standard Lena portrait
(Figure~\ref{fig:private:fourier}(a))
using two finite fields of different characteristics: $\GF(13)$ and $\GF(73)$.
The obtained watermarked images are shown in Figure~\ref{fig:private:fourier}(c--d),
respectively.
The calculation over $\GF(73)$ furnished an image
whose degradation is visually perceptible.

Indeed, the described experiment qualitatively demonstrates
a limitation in Tamori-Aoki-Yamamoto scheme.
Because visually transparent watermarked images are not always achievable,
the choice of finite fields becomes restricted to
small values of $p$.

\section{A Signature System}
\label{sectionsignature}

Apart from the private watermarking mode of operating,
the
described scheme can be interpreted from a different, new perspective.
The present study proposes
that the same system can be also considered as
a method for signature data generation.

The watermark embedding process is now
understood as a signature generation operation.
The output signature~$\mathbf{S}$ is simply defined as
\begin{equation}
\mathbf{S} \triangleq \mathbf{D}',
\end{equation}
where the image~$\mathbf{D}'$ is obtained according to
Equation~\ref{eq-watermarked}.
Once obtained, the signature data can be made publicly available.

In this context,
the original watermark extraction method becomes an authentication operation
performed on raw data~$\mathbf{D}$.
The input data to be verified is~$\mathbf{D}$,
which is owned by the user.
Therefore,
the associated signature data,
which can be retrieved from a trusted source,
and
the raw image can be submitted to
the discussed method.
An unaltered extracted watermark
is an evidence of
the integrity and authenticity of~$\mathbf{D}$.
In the case of a tampering attack,
either on~$\mathbf{D}$ or on~$\mathbf{S}$,
the extracted watermark
can indicate the data corruption
location.

It is worthwhile to emphasize that the aim is not to produce perceptually transparent
watermarked images.
Therefore,
this approach completely diminishes the sensitivity to the value of
the field characteristic,
as experienced in the original methodology.

Again because the
NTT transform domain lacks a physical meaning,
even an alteration in the least significant bit of a signal
can render a totally different NTT spectrum.
This is a desirable property.
No matter how subtle,
tampering
can
produce
strikingly
noticeable differences in the recovered watermark.
Consequently,
this approach can
verify the authenticity and integrity of raw digital data,
providing means to spatially locate alteration regions.

\section{Computational Results and Discussion}
\label{sectioncomputational}

In this section,
some computational experiments
are performed to
validate
the proposed
methods.
Selected standard images
available at
the University of Southern California
Signal and Image Processing Institute (USC-SIPI) Image Database~\cite{uscsipi}
were submitted to the
suggested watermarking scheme.
Private watermarking and signature generation
functionalities of the method
were explored,
with the application of
the two-dimensional finite field cosine and Hartley transforms.

\begin{figure}
\centering
\includegraphics{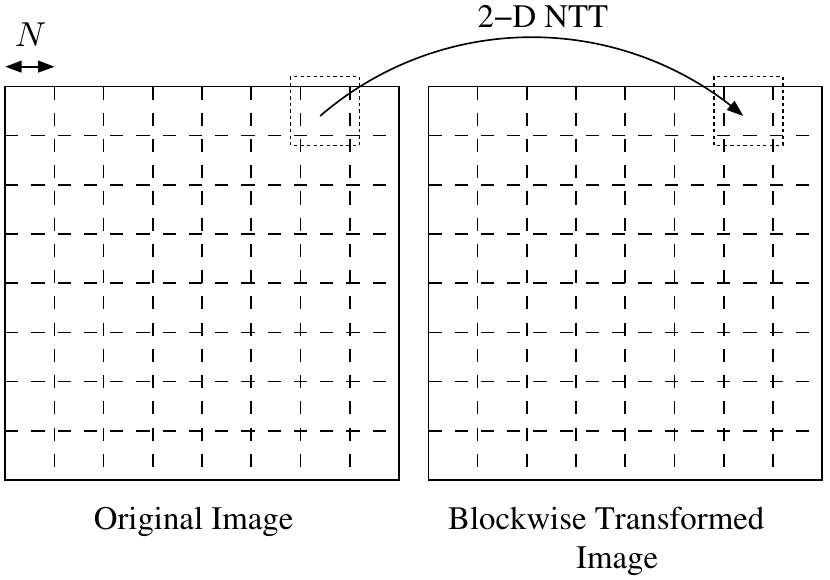}
\caption{A blockwise two-dimensional transform application.
The size of the the dashed blocks matches the two-dimensional transform blocksize $N\times N$.}
\label{blockwise}
\end{figure}

Generally the dimensions of a practical number theoretic transformation matrix $N\times N$
are smaller than usual image sizes.
Typical values of the transform size $N$ are given in Table~\ref{tab:zetas:ffct} and~\ref{tab:zetas:ffht}.
Therefore,
the required two-dimensional transforms are obtained
conforming to a blockwise computation.
This approach simply consists of
decomposing a given host image into adjacent,
equal sized, nonoverlapping blocks.
These constituent blocks must match the dimensions
of the considered two-dimensional transformation matrix.
After that,
each block is
transformed
and
the resulting 
\mbox{2-D}~spectral 
blocks are reassembled,
which results in a
blockwise transformed image.
Figure~\ref{blockwise} illustrates the process.
Accordingly,
the discussed watermarking schemes are applied blockwise 
as well.

\begin{figure}
\centering
\subfigure[$512\times 512$]{\includegraphics[width=0.3\linewidth,height=0.3\linewidth]{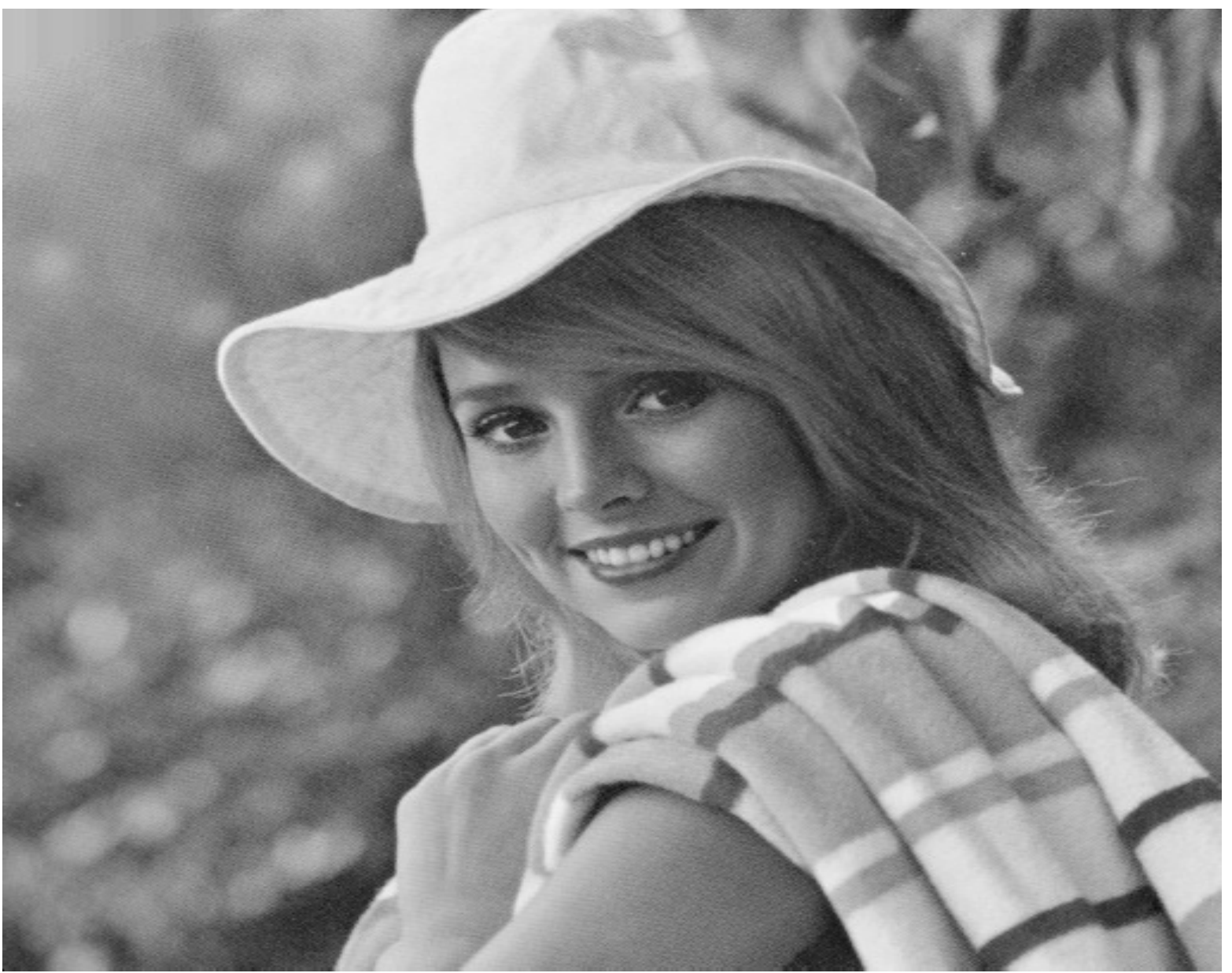} }
\subfigure[$512\times 512$]{\includegraphics[width=0.3\linewidth,height=0.3\linewidth]{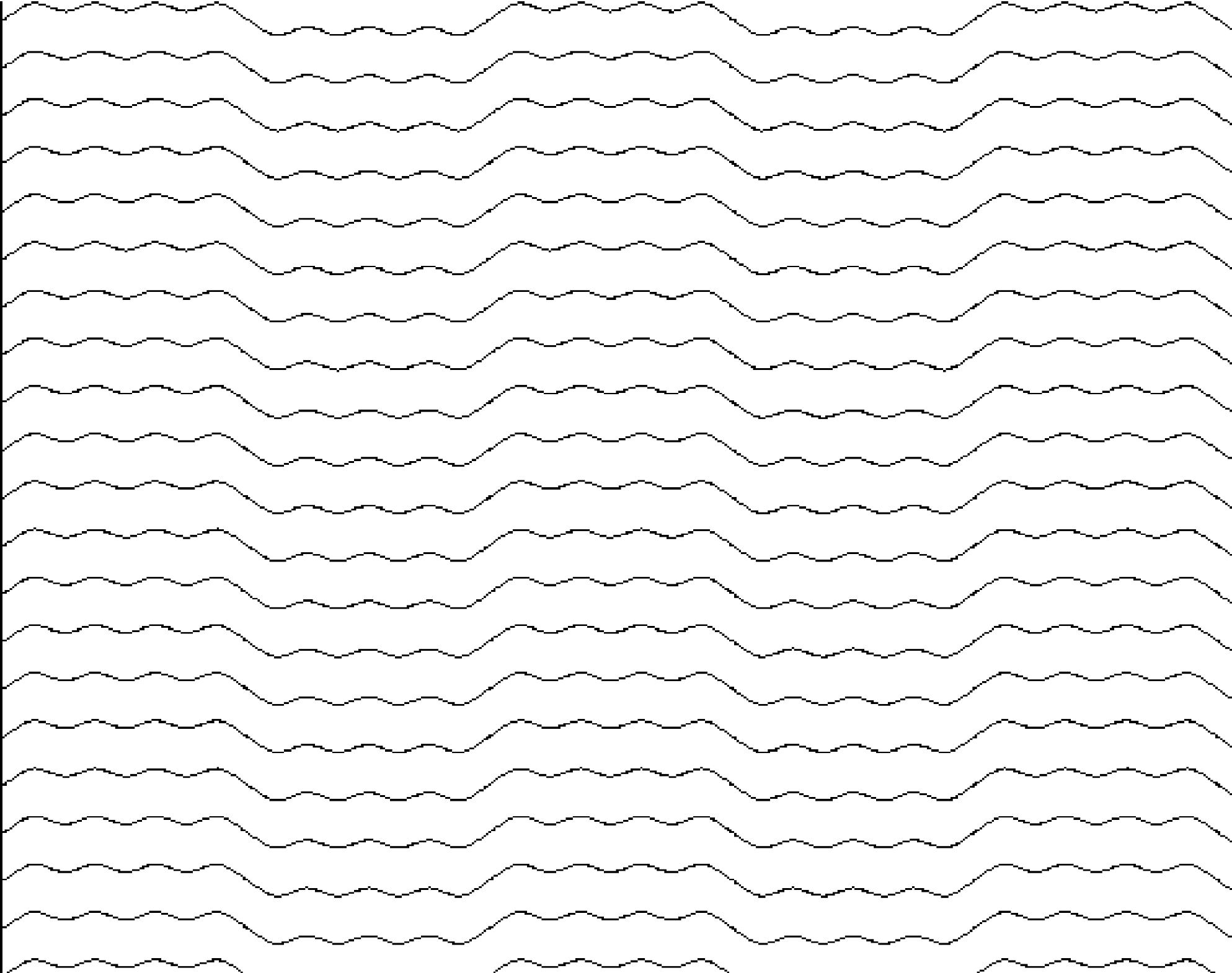} }
\subfigure[$512\times 512$]{\includegraphics[width=0.3\linewidth,height=0.3\linewidth]{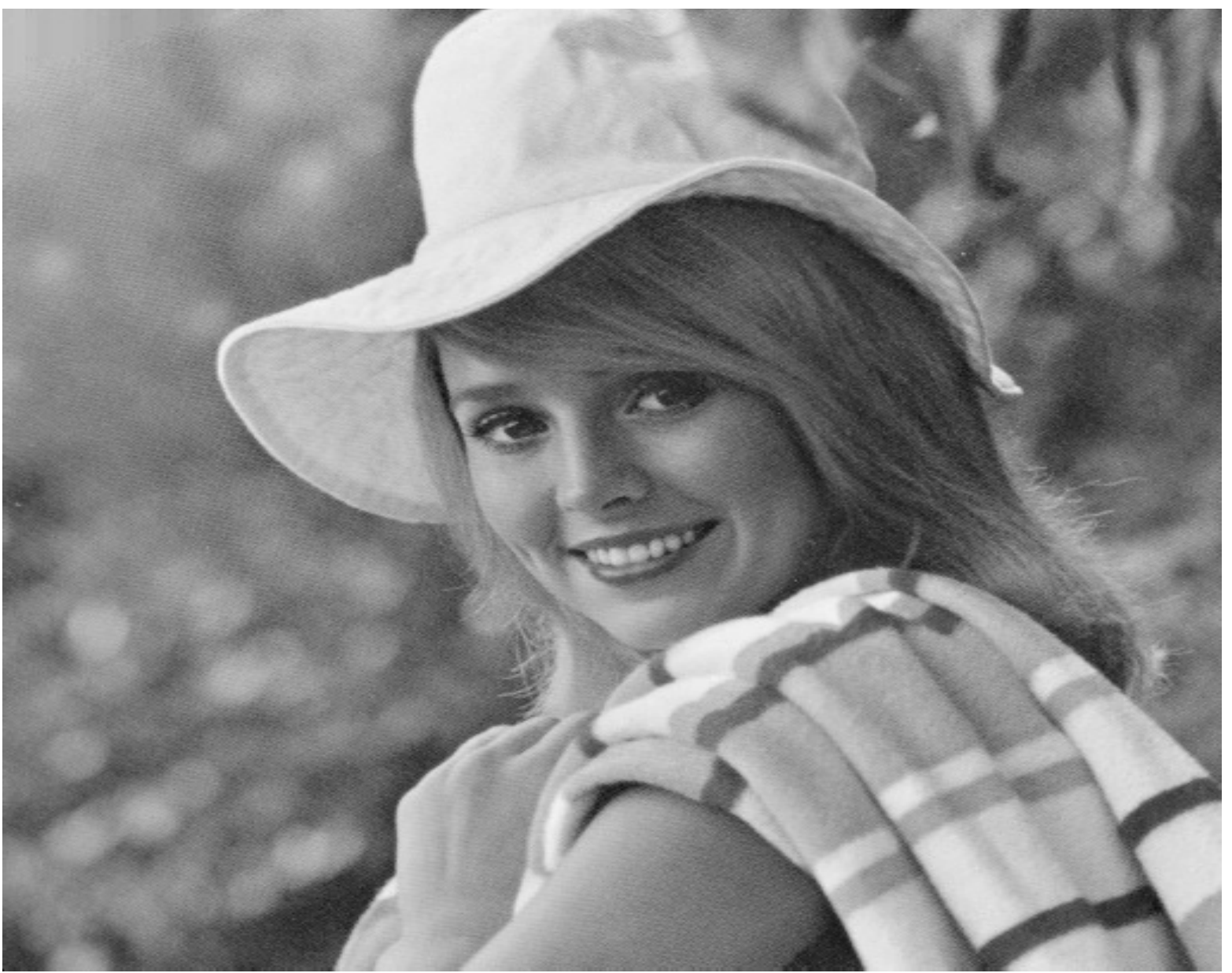} } \\
\subfigure[$512\times 512$]{\includegraphics[width=0.45\linewidth,height=0.45\linewidth]{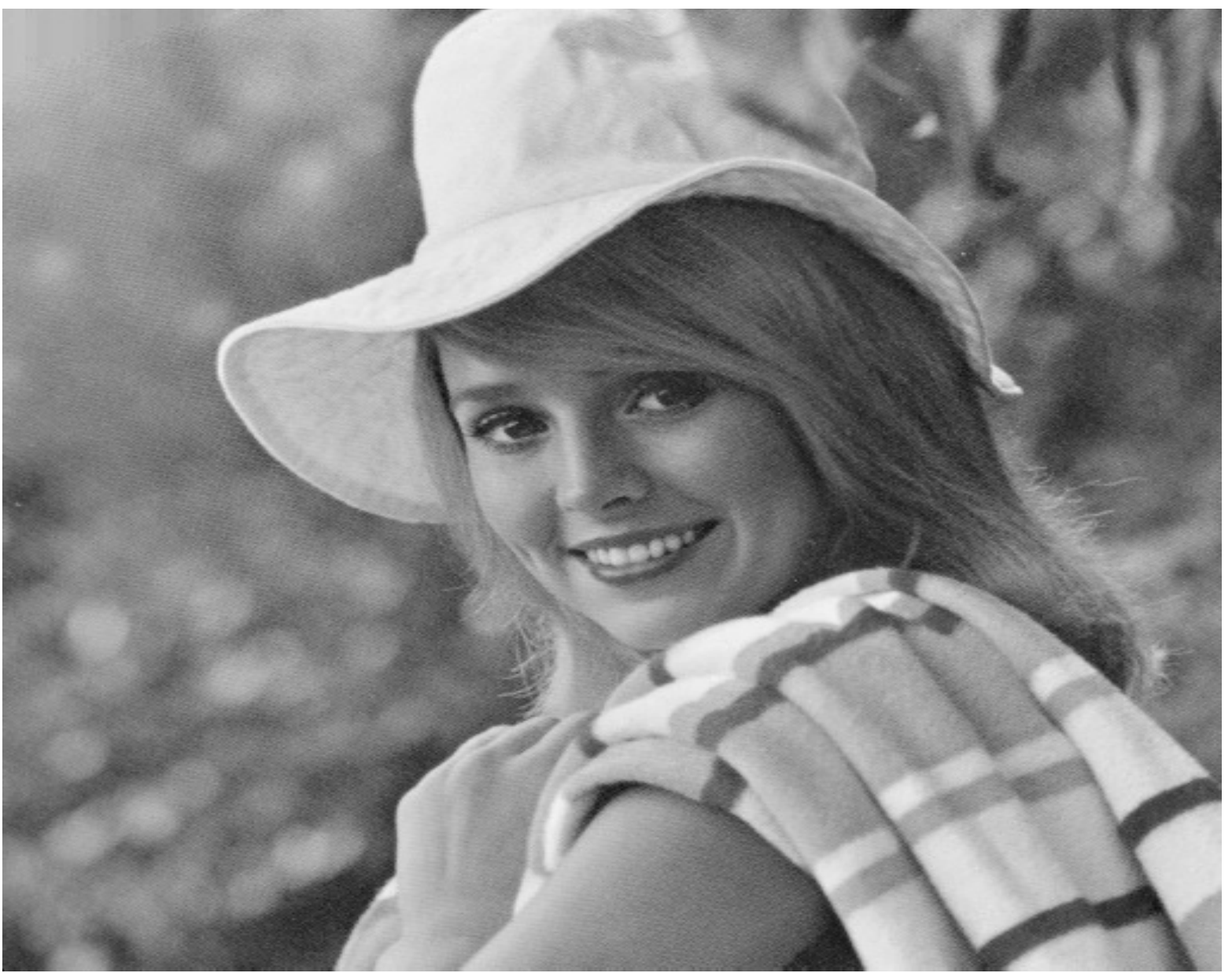} }
\subfigure[$512\times 512$]{\includegraphics[width=0.45\linewidth,height=0.45\linewidth]{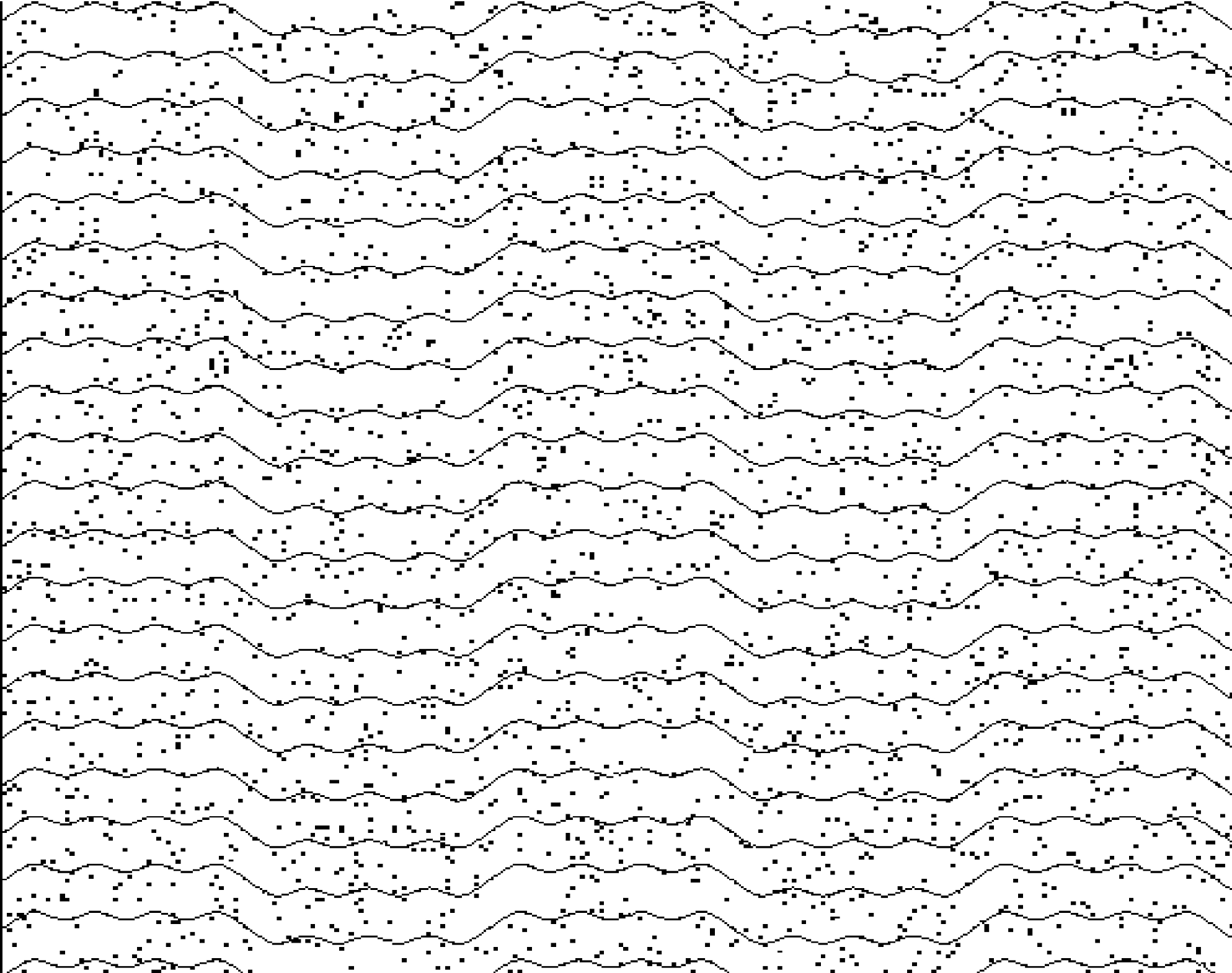} }
\caption{Private watermarking using the FFCT.
(a)~Original image \texttt{elaine.512.tiff} (Elaine),
(b)~watermark (negative image for better visualization),
(c)~watermarked image,
(d)~tampered image with a Bernoulli distributed noise of $10^{-2}$ error probability,
(e)~extracted watermark with clear location of tampering (negative image for better visualization).
Calculations were performed over $\GF(7)$ with $\zeta = 2+2j$, transform size is 2.}
\label{fig:private:cosine}
\end{figure}

Considering the FFCT,
a raw image (Figure~\ref{fig:private:cosine}(a))
was processed and furnished the watermarked data displayed in Figure~\ref{fig:private:cosine}(c).
Figure~\ref{fig:private:cosine}(d)
shows the original image after the introduction of random artifacts on its pixel values.
The modification consisted of
incrementing the image pixels by one
with probability $10^{-2}$.
A wavy pattern was selected as the watermark (Figure~\ref{fig:private:cosine}(b)) and
an FFCT over $\GF(7)$ with blocksize of two was chosen.
Over $\GF(7)$,
a possible choice is $\zeta=2+2j$,
which induces an NTT of length equal to 2.
This small blocksize was able to sharply indicate tampering locations (Figure~\ref{fig:private:cosine}(e)).

\begin{figure}
\centering
\subfigure[$256 \times 256$]{\includegraphics[width=0.3\linewidth,height=0.3\linewidth]{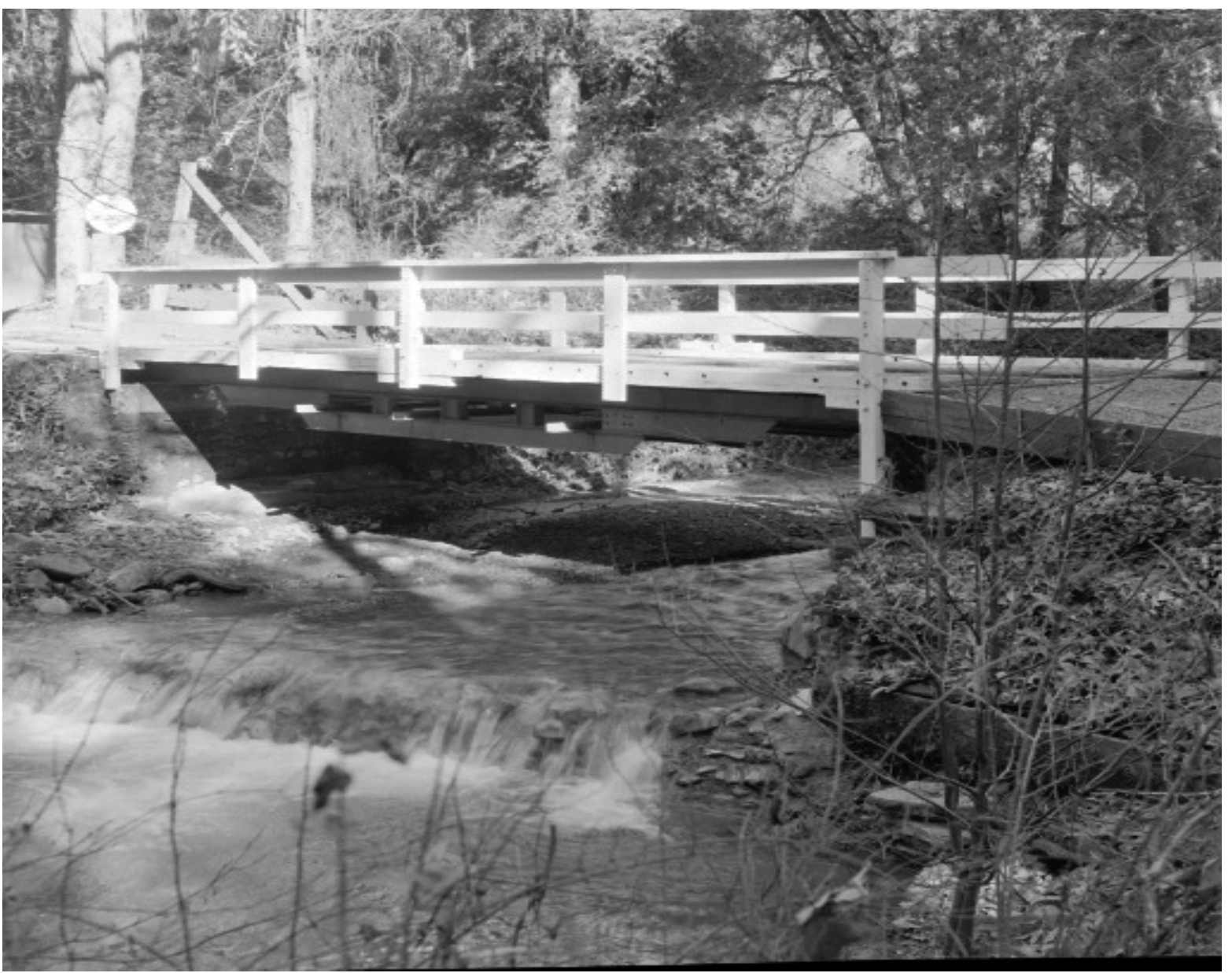} }
\subfigure[$64 \times 64$]{\includegraphics[width=0.3\linewidth,height=0.3\linewidth]{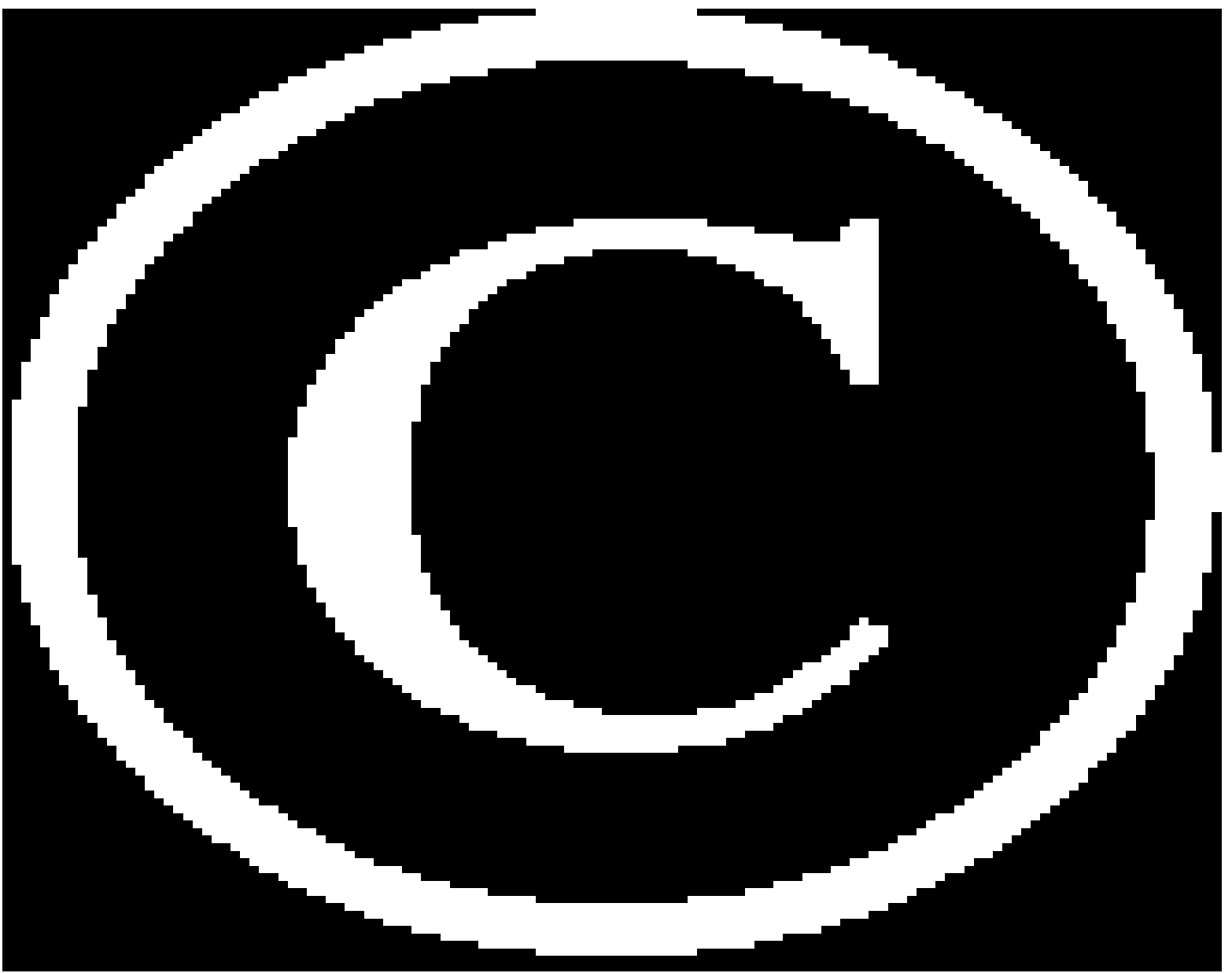} }
\subfigure[$256 \times 256$]{\includegraphics[width=0.3\linewidth,height=0.3\linewidth]{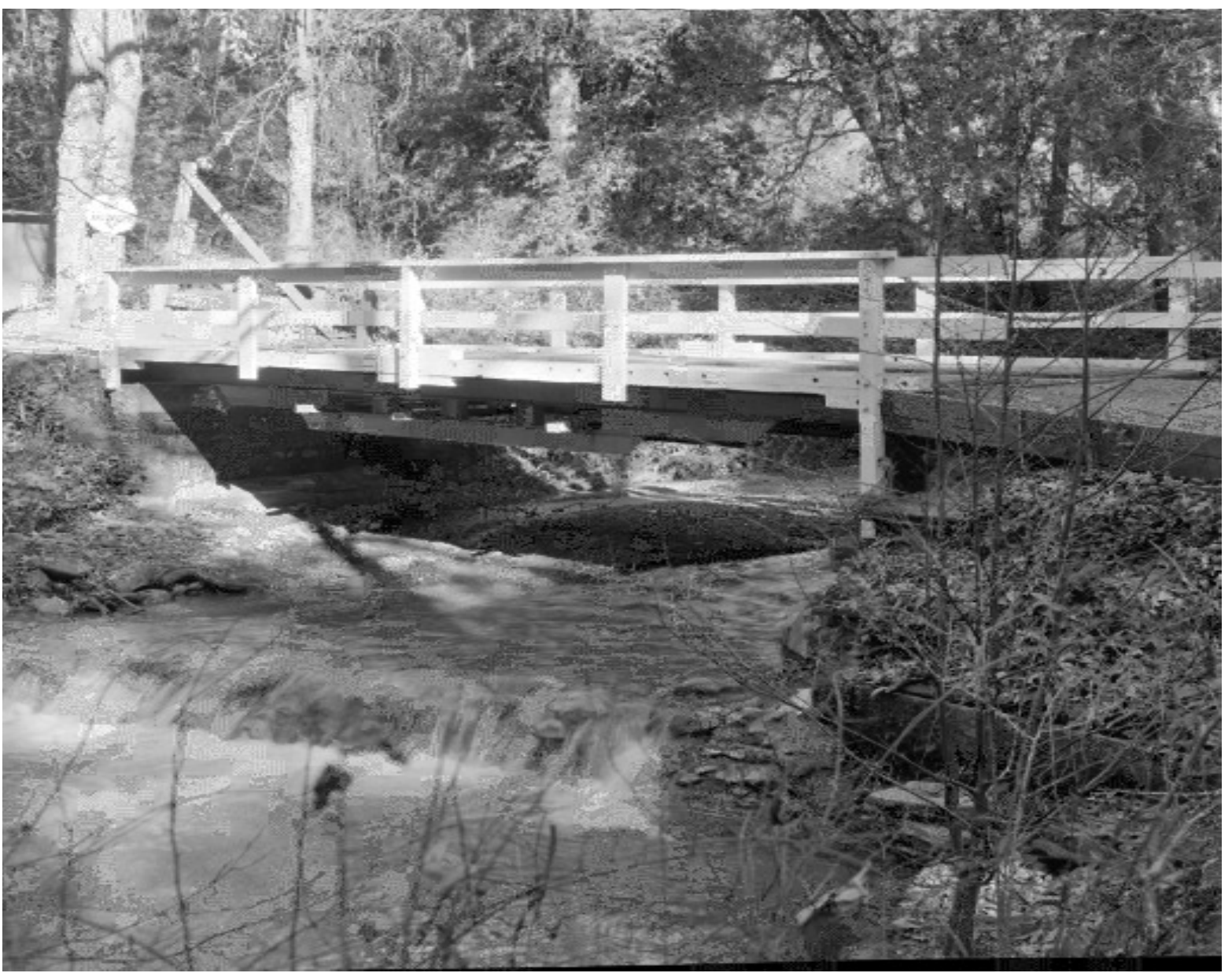} } \\
\subfigure[$256 \times 256$]{\includegraphics[width=0.45\linewidth,height=0.45\linewidth]{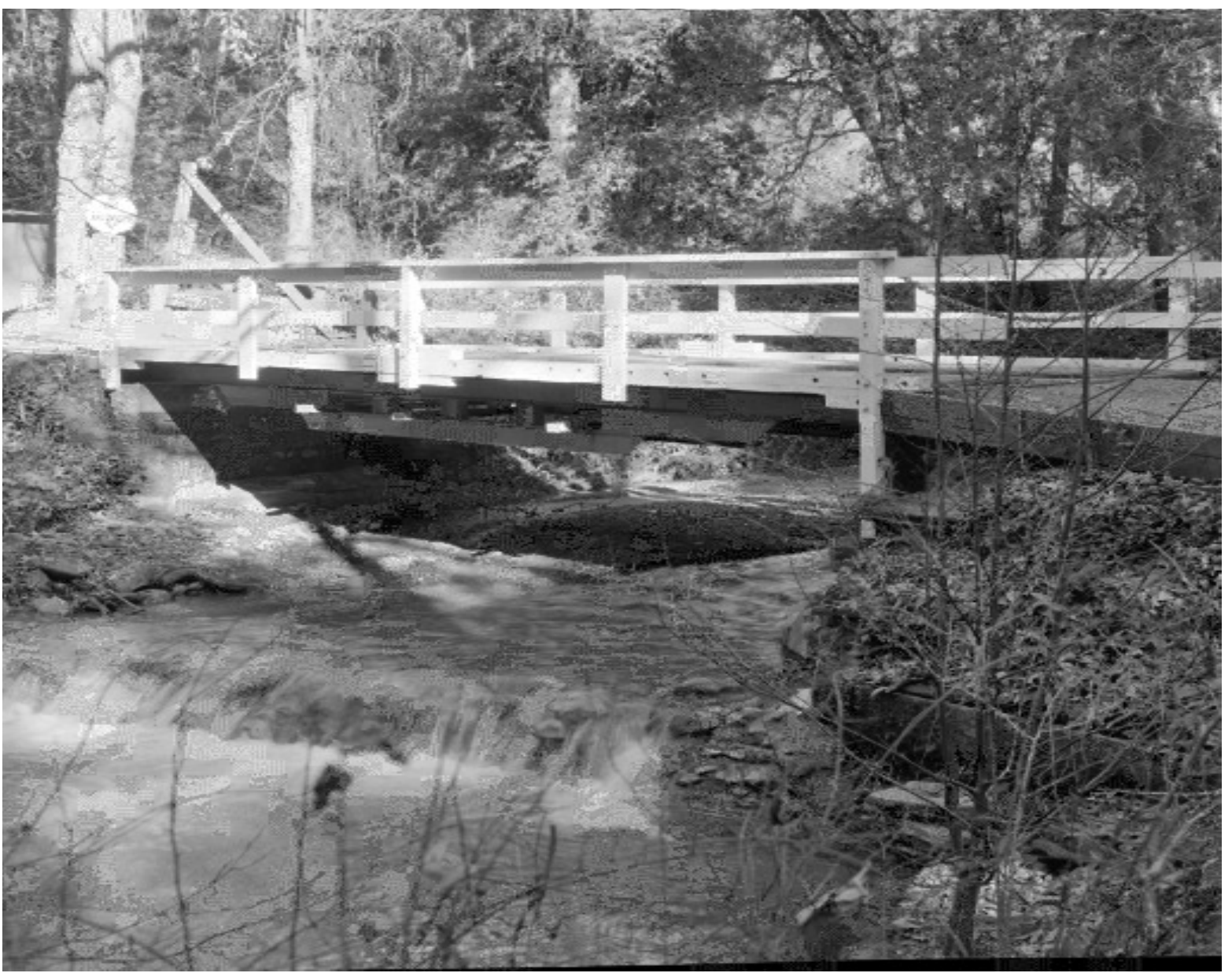} }
\subfigure[$256 \times 256$]{\includegraphics[width=0.45\linewidth,height=0.45\linewidth]{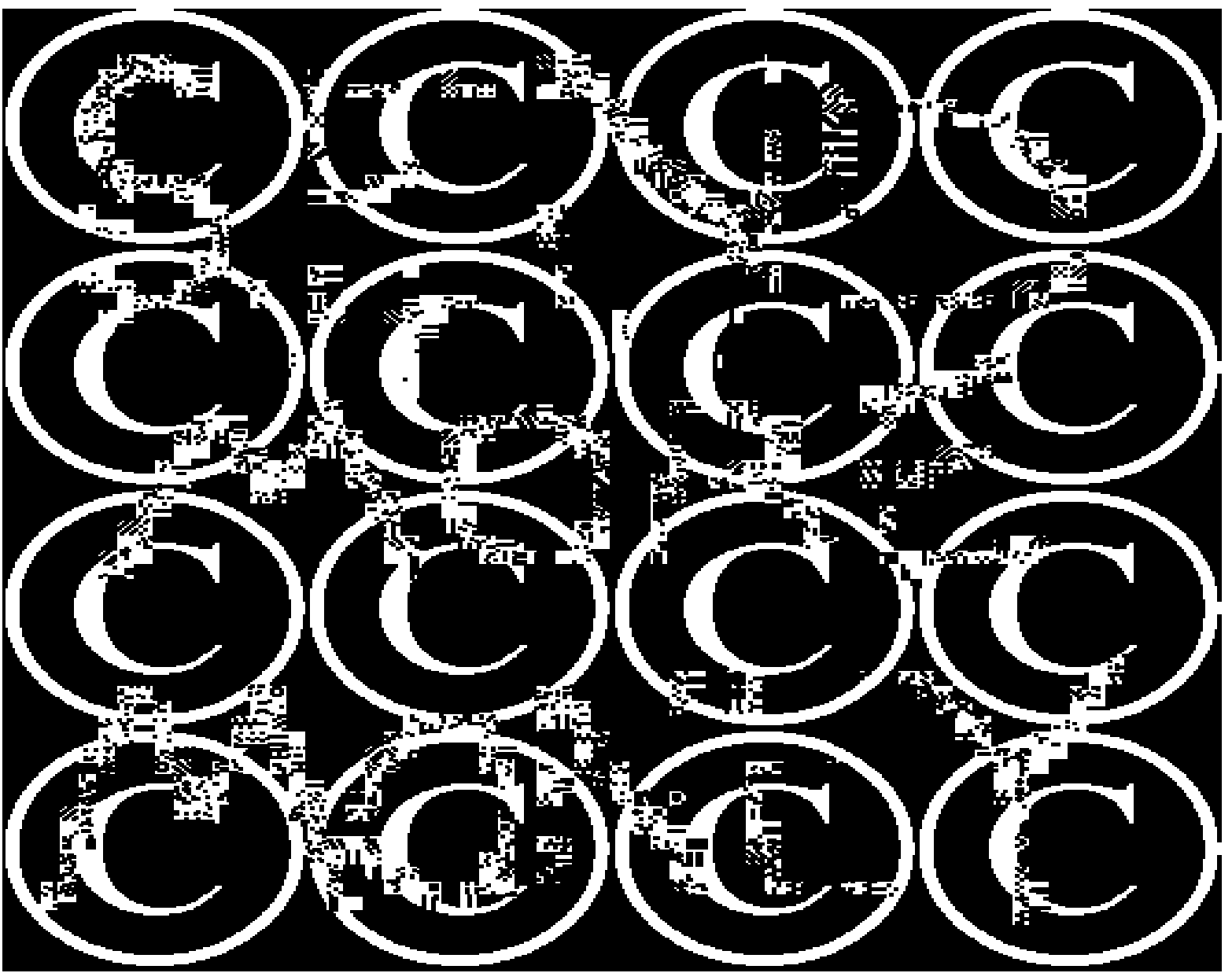} }
\caption{Private watermarking using the FFHT.
(a)~Original image \texttt{5210.tiff} (Stream and bridge),
(b)~watermark,
(c)~watermarked image,
(d)~tampered image with the additive insertion of another image,
(e)~extracted watermark with a hidden message (\textsc{send more money}).
Calculations were performed over $\GF(31)$ with $\zeta = 4+4j$, transform size is 8.}
\label{fig:private:hartley}
\end{figure}

Similar results were obtained with the aid of the FFHT.
Using an 8-point \mbox{2-D}~FFHT over $\GF(31)$ ($\zeta = 4+4j$),
a modification introduced into a raw image (Figure~\ref{fig:private:hartley}(a))
could be detected and assessed.
Differently from the previous experiment,
a smaller watermark was utilized (Figure~\ref{fig:private:hartley}(b)).
After the calculations,
the recovered watermark image (Figure~\ref{fig:private:hartley}(e))
clearly discloses the tampering
and
shows the interfered regions.
Actually,
the use of a small transform blocklength made
possible to provide information
on the nature of the modification.
In this case, a small text had been written on
the original image.
If a larger transform were employed,
this capability would be reduced,
because the recovered watermark would present
the tampered areas in
larger blocks.
Ultimately the accuracy of
the tampering location would be mitigated.

\begin{figure}
\centering
\subfigure[$256 \times 256$]{\includegraphics[width=0.3\linewidth,height=0.3\linewidth]{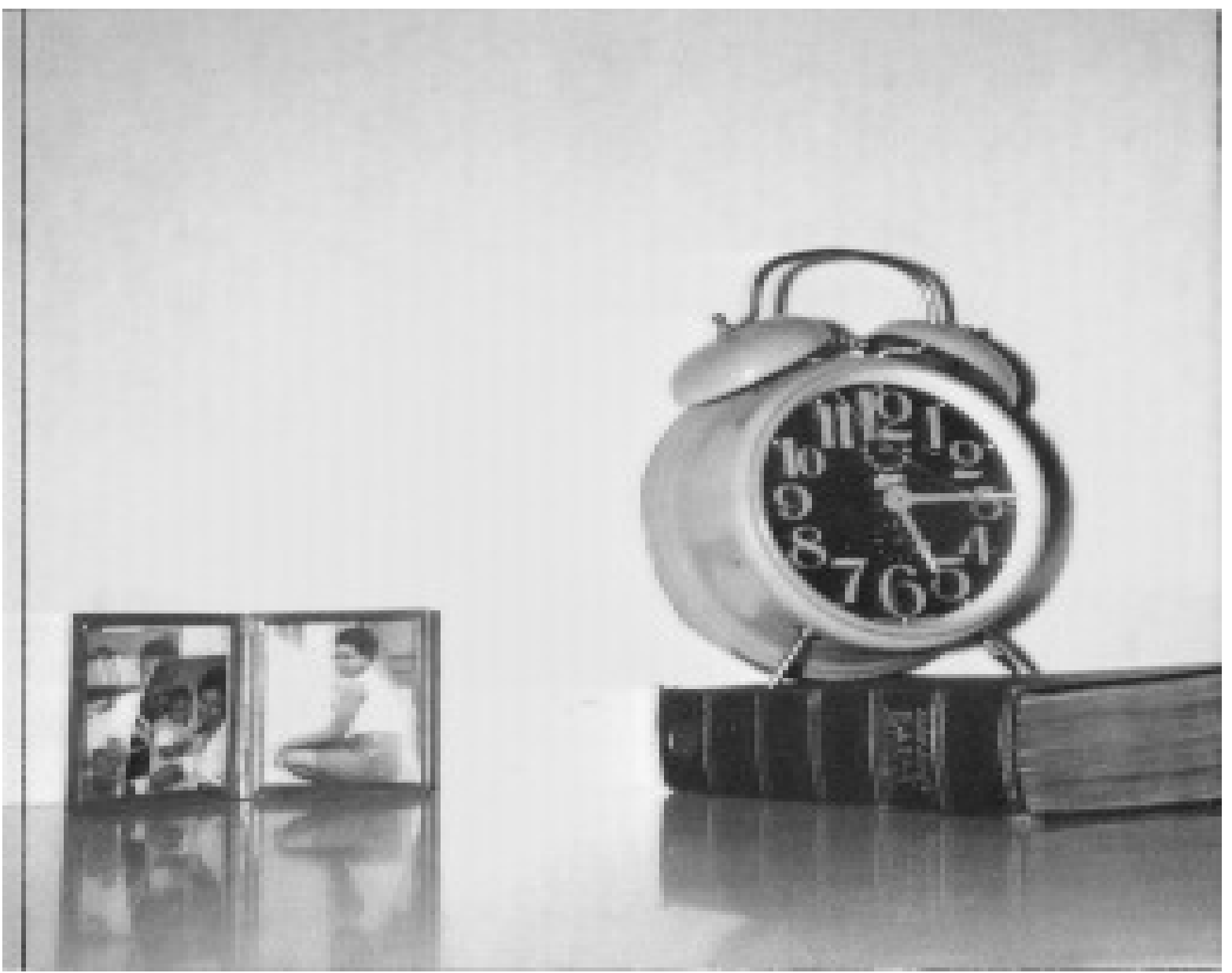} }
\subfigure[$64 \times 64$]{\includegraphics[width=0.3\linewidth,height=0.3\linewidth]{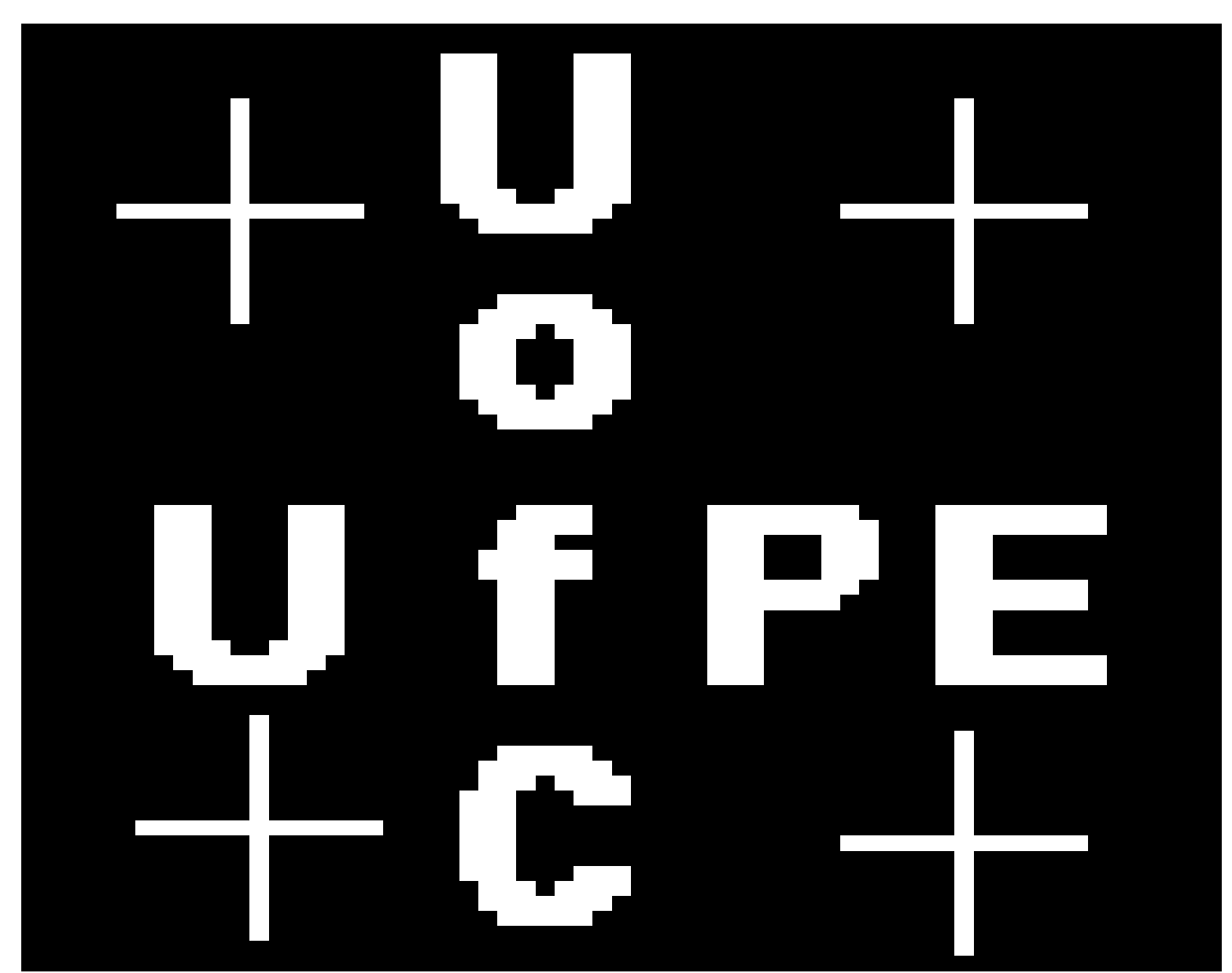} }
\subfigure[$256 \times 256$]{\includegraphics[width=0.3\linewidth,height=0.3\linewidth]{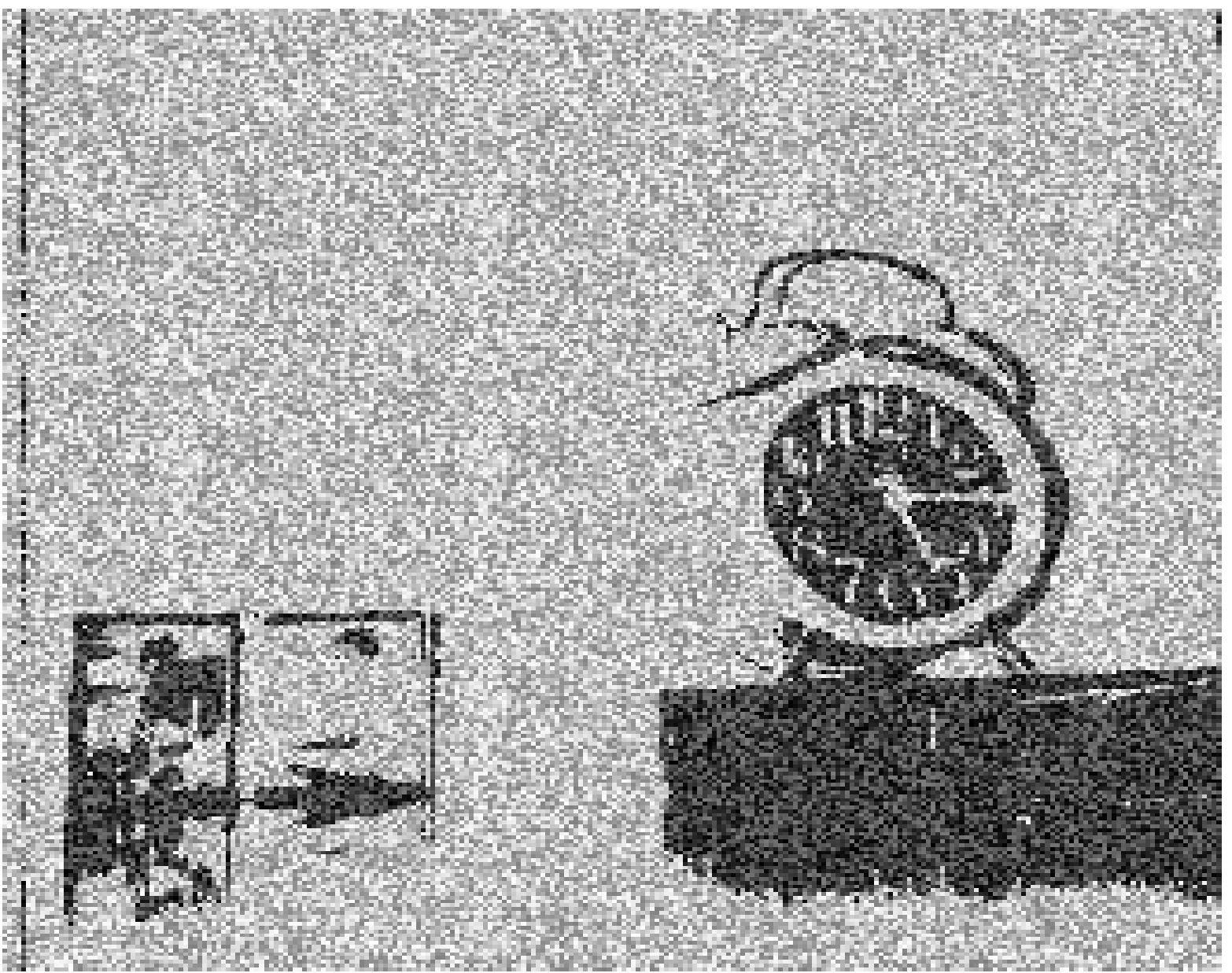} } \\
\subfigure[$256 \times 256$]{\includegraphics[width=0.45\linewidth,height=0.45\linewidth]{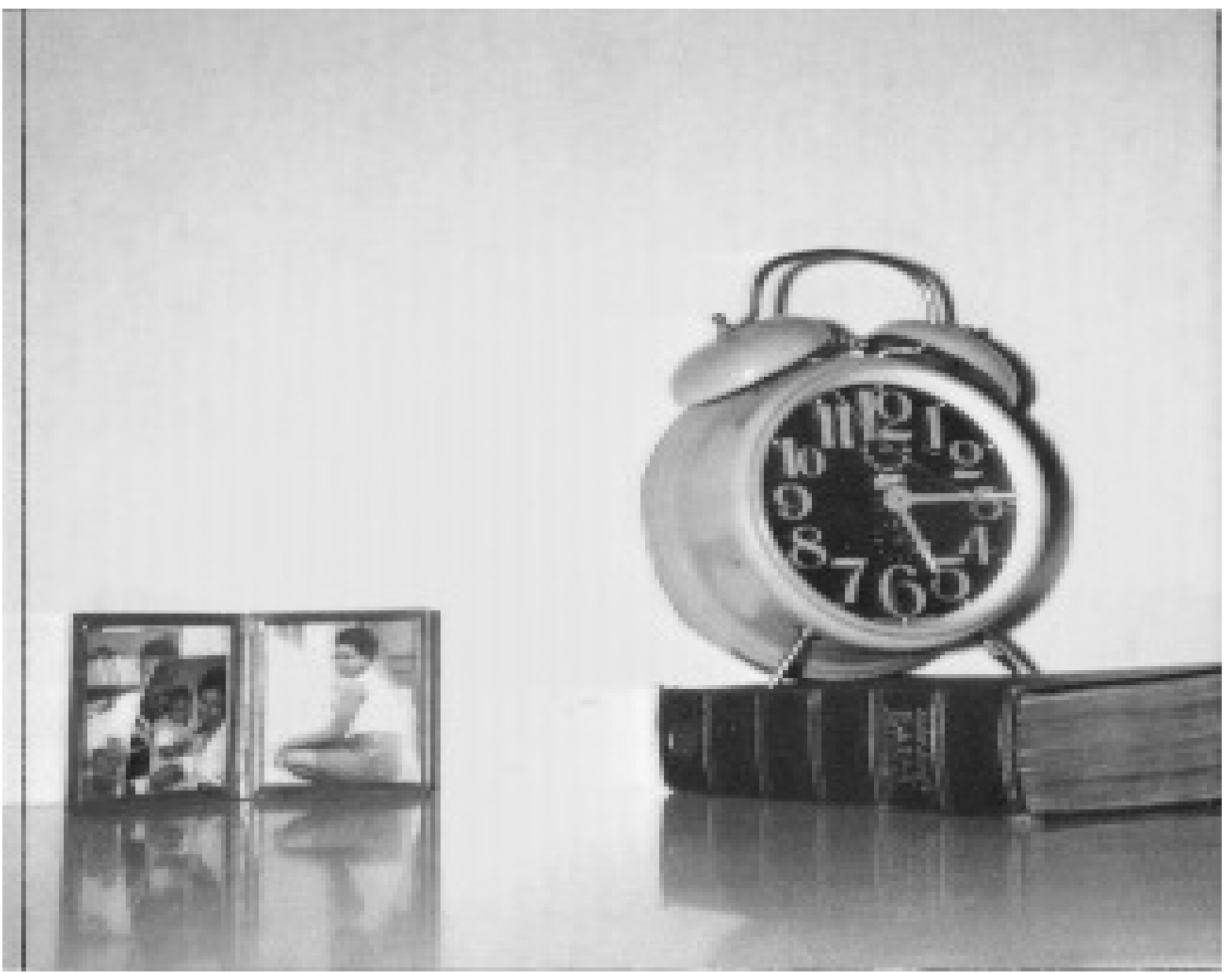} }
\subfigure[$256 \times 256$]{\includegraphics[width=0.45\linewidth,height=0.45\linewidth]{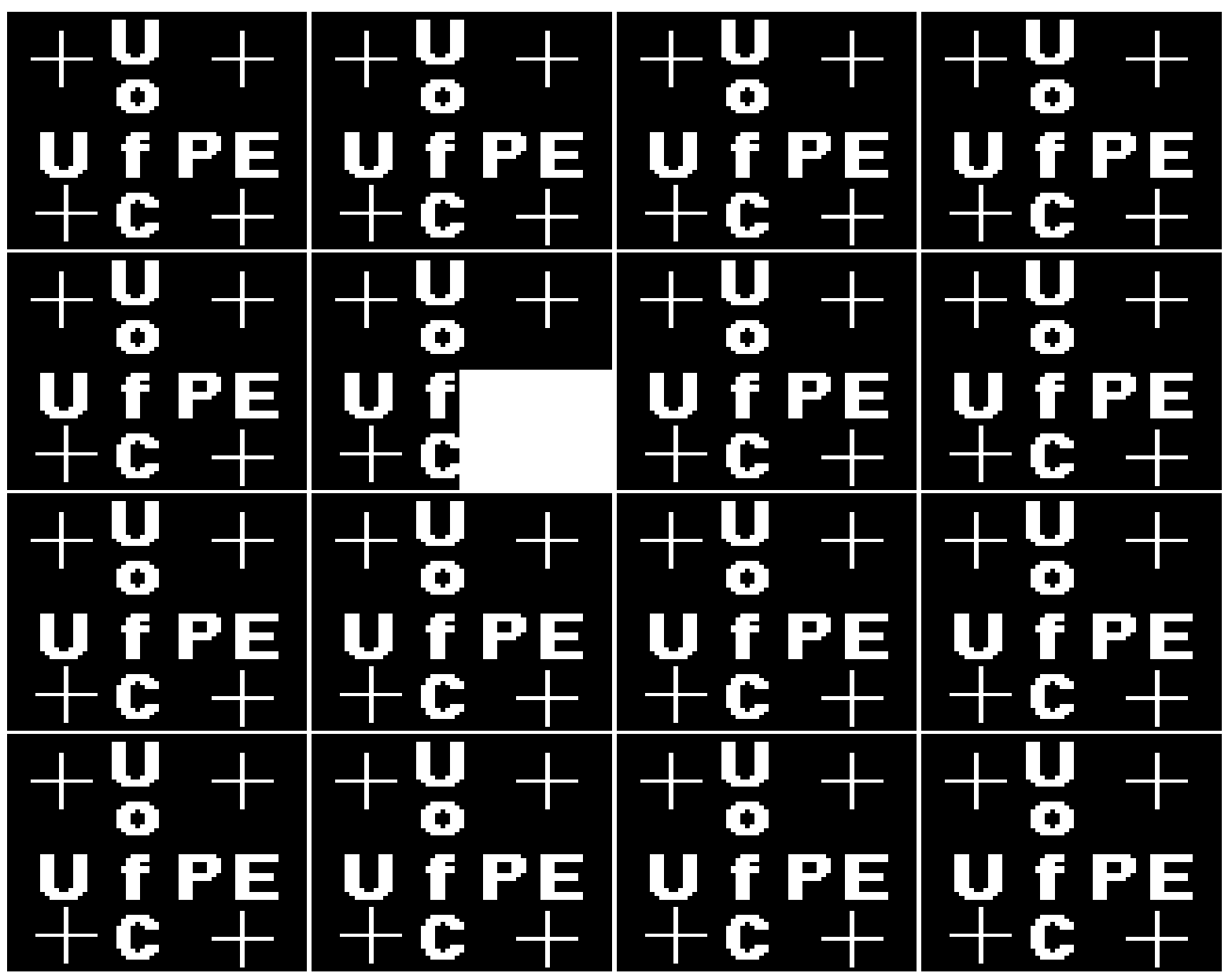} }
\caption{Signature data generation using the FFCT.
(a)~Original image \texttt{5112.tiff} (Clock),
(b)~watermark,
(c)~signature image (publicly available),
(d)~tampered image with the pixel at position $(100,100)$ incremented by one,
(e)~extracted watermark.
Observe the fragility of the signature embedding: a single bit alteration made a whole block
of recovered data  to be distorted.
Calculations were performed over $\GF(127)$ with $\zeta = 2+39j$, transform size is 32.}
\label{fig:public:cosine}
\end{figure}

\begin{figure}
\centering
\subfigure[$512 \times 512$]{\includegraphics[width=0.3\linewidth,height=0.3\linewidth]{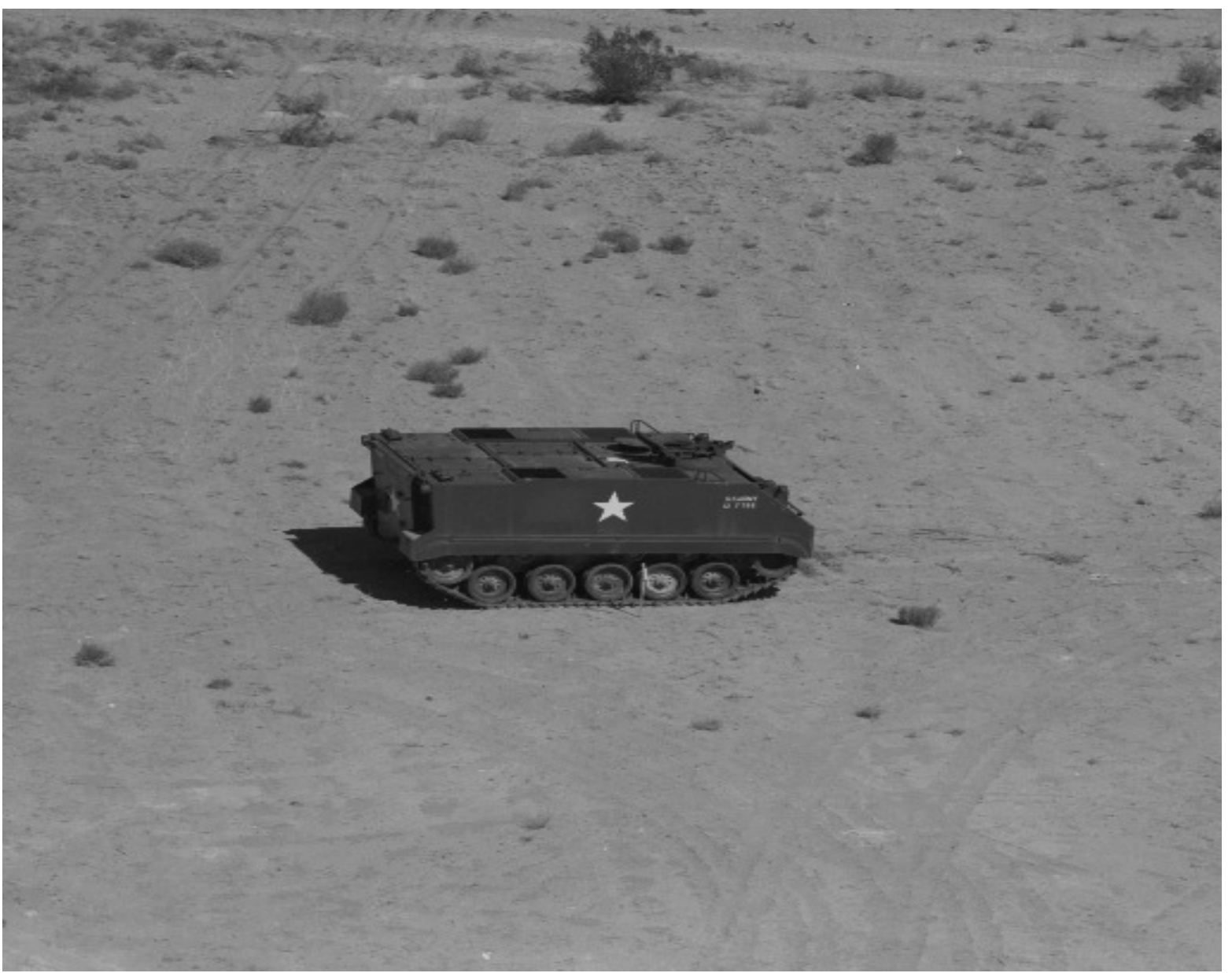} }
\subfigure[$64 \times 64$]{\includegraphics[width=0.3\linewidth,height=0.3\linewidth]{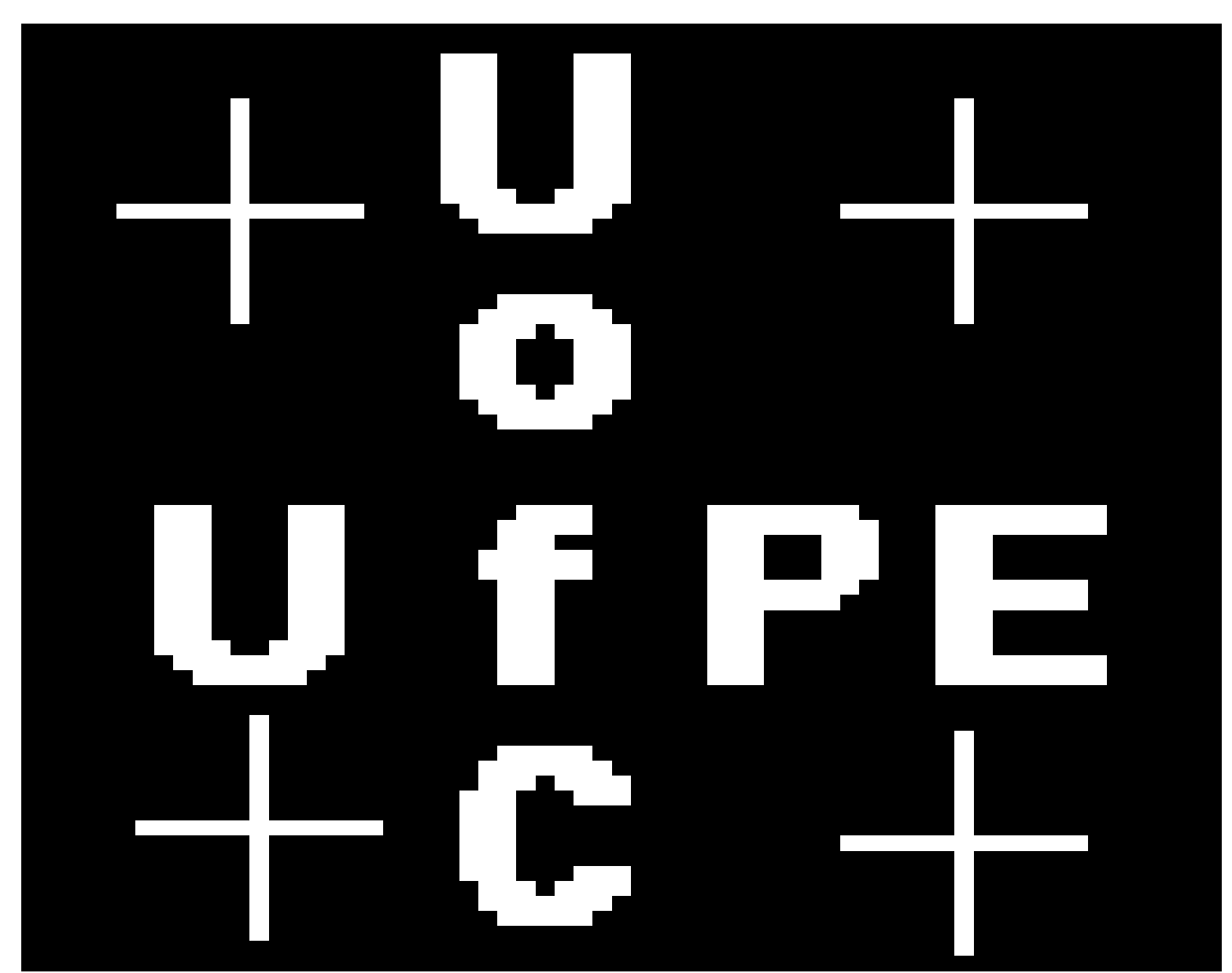} }
\subfigure[$512 \times 512$]{\includegraphics[width=0.3\linewidth,height=0.3\linewidth]{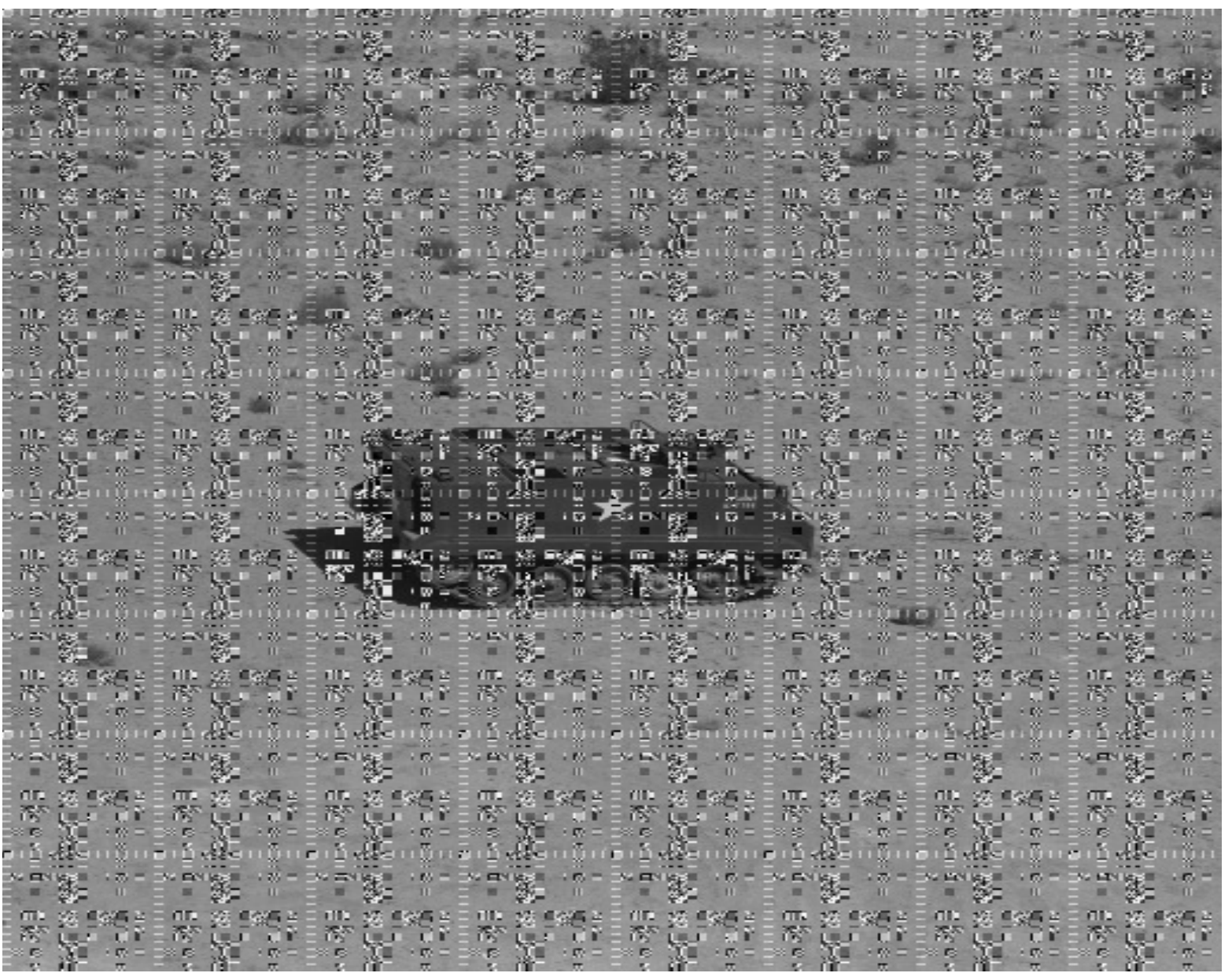} } \\
\subfigure[$512 \times 512$]{\includegraphics[width=0.45\linewidth,height=0.45\linewidth]{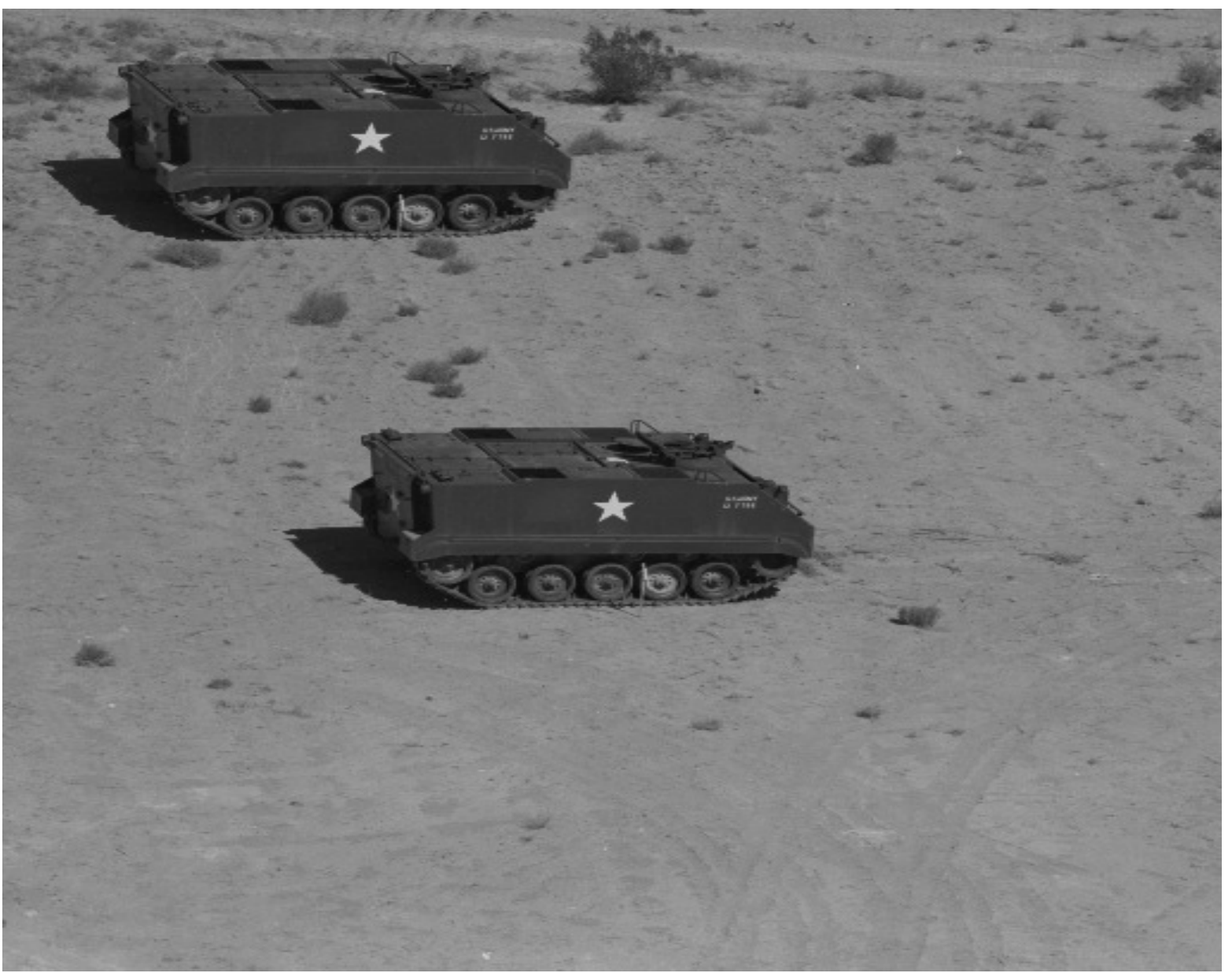} }
\subfigure[$512 \times 512$]{\includegraphics[width=0.45\linewidth,height=0.45\linewidth]{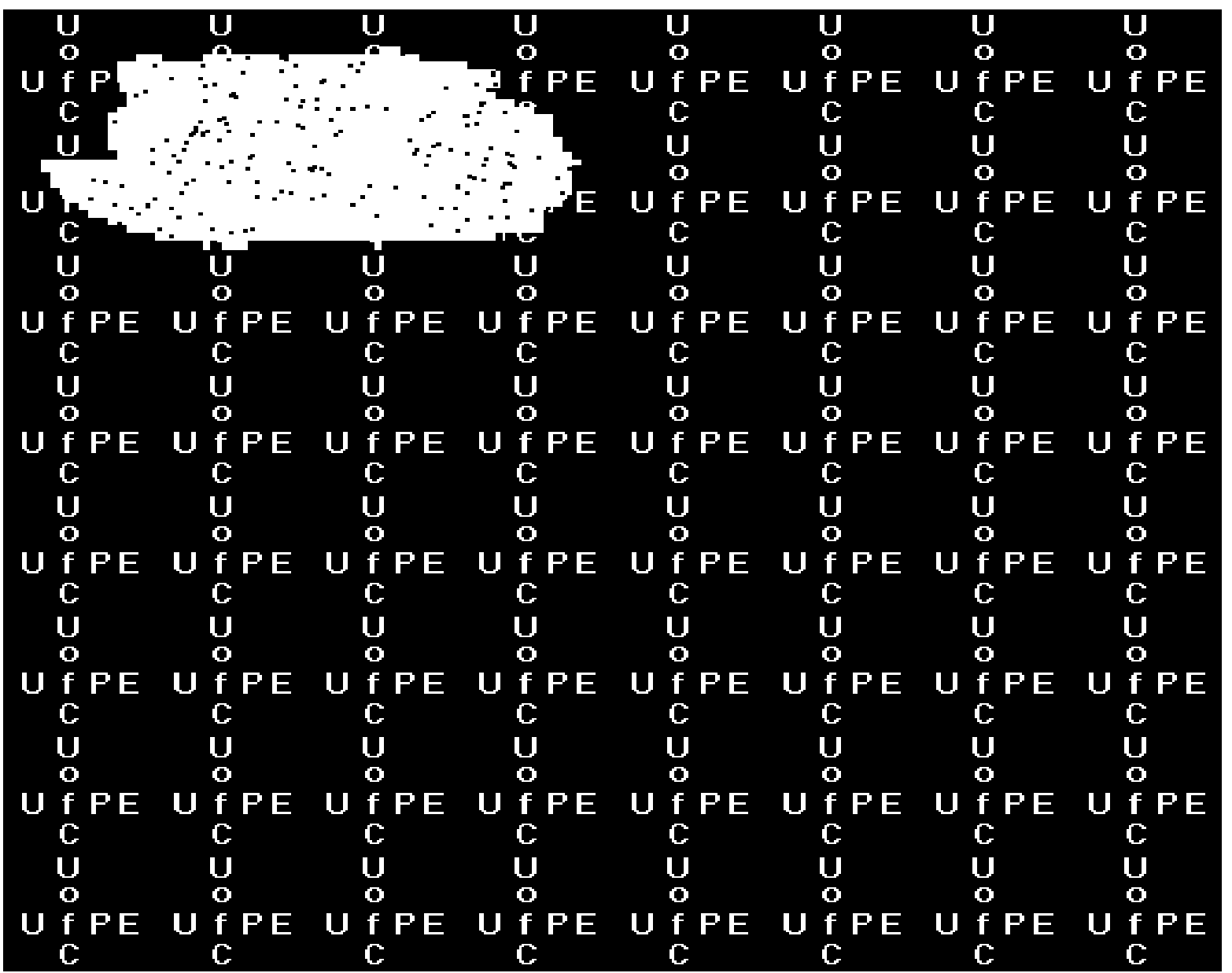} }
\caption{Signature data generation using the FFHT.
(a)~Original image \texttt{7108.tiff} (APC),
(b)~watermark,
(c)~signature image (publicly available),
(d)~tampered image,
(e)~extracted watermark.
Calculations were performed over $\GF(251)$ with $\zeta = j$, transform size is 4.}
\label{fig:public:hartley}
\end{figure}

Figures~\ref{fig:public:cosine} and~\ref{fig:public:hartley}
illustrate
the signature operation proposed in the current work.
An original image (Figure~\ref{fig:public:cosine}(a))
was processed with the inclusion of a watermark
(Figure~\ref{fig:public:cosine}(b)).
Consequently, the signature image data depicted
in Figure~\ref{fig:public:cosine}(c)
was obtained.
Being the calculations over $\GF(127)$,
a visually low-quality image was attained.
However, this is not important, because the
Figure~\ref{fig:public:cosine}(c) is
only intended to be
used as
signature data.
Only its mathematical properties are relevant.
A tampering consisting of incrementing a single pixel at position $(100,100)$
was applied to the original image,
and Figure~\ref{fig:public:cosine}(d) was derived.
The output of the authentication system is
shown in Figure~\ref{fig:public:cosine}(e).
As the chosen $\zeta$ for this calculations was
$2+39j$,
the transform blocksize was 32.
Therefore,
the precision of tampering positioning is restricted to $32 \times 32$ pixel regions.

For an improved tampering location accuracy,
one may consider shorter transforms,
as
in
the experiment depicted in Figure~\ref{fig:public:hartley}.
After signature generation in $\GF(251)$ using the FFHT
with~$\zeta=j$ (Figure~\ref{fig:public:hartley}(c)),
the host data (Figure~\ref{fig:public:hartley}(a)) was maliciously tampered.
Figure~\ref{fig:public:hartley}(d) shows that a forged image is introduced.
This specific~$\zeta$ implies a 4-point transform,
which provides a potentially adequate tampering location accuracy.
The recovered watermark clearly discriminates the modification site (Figure~\ref{fig:public:hartley}(e)).

\begin{figure}
\centering
\includegraphics[width=0.8\linewidth]{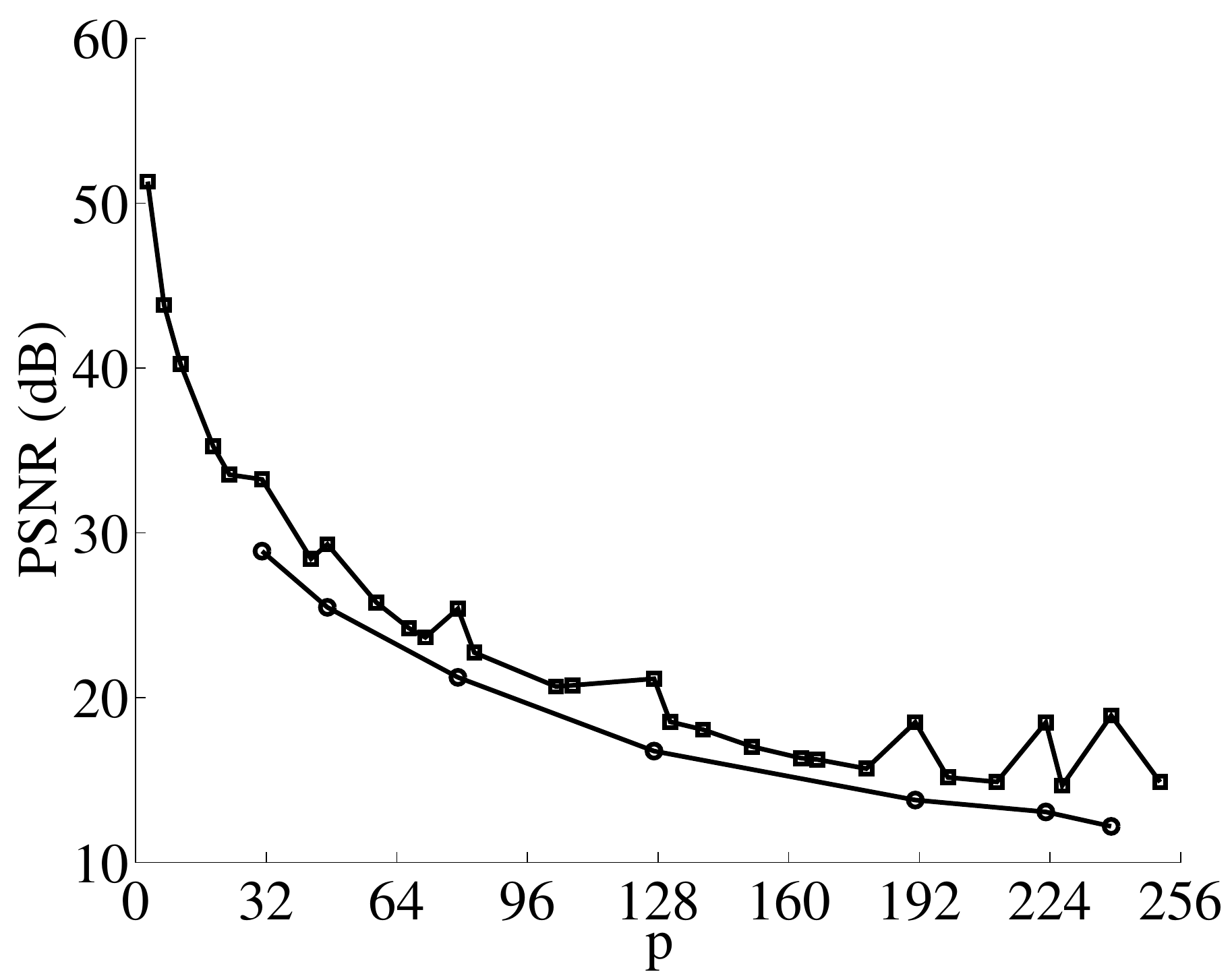}
\caption{Quality degradation of watermarked images generated by the discussed method as
a function of the finite field characteristic.
The PSNR was employed to assess the image degradation.
The Lena portrait was watermarked with 4-point \mbox{2-D}~FFCT (\mbox{---$\circ$---}) and FFHT (\mbox{---$\Box$---})
in all pertinent Galois fields $\GF(p)$, for $3\leq p \leq 251$.}
\label{fig:psnr}
\end{figure}

Given a fixed subject image (Lena portrait),
watermarked/signature images were generated via the method discussed herein,
for all possible finite fields with characteristic ranging from 3 to 251.
Both FFCT and FFHT were considered and~$\zeta$ was chosen in such a way
that 4-point \mbox{2-D}~transformations were guaranteed.
Subsequently,
the peak signal-to-noise ratio (PSNR)
was
adopted as a
measure of the image deterioration after the watermarking process.
For small values of~$p$, visually imperceptible watermarks can be produced,
since the resulting images present high PSNR values.
However, for large values of~$p$, the image quality becomes unacceptable for invisible watermarking purposes.
Despite of that, for signature data generation,
obtaining low PSNR values becomes less relevant.
Quantitative data
concerning the image degradation
are summarized in Figure~\ref{fig:psnr}.

\section{Conclusion}
\label{sectionconclusion}

The utilization of finite fields as an avenue for fragile watermarking was
explored.
The watermarking scheme suggested in~\cite{tamori2002fragile}
was amplified and generalized by
the introduction of finite field trigonometrical transforms,
such as the finite field cosine and Hartley transforms.
The new proposed methodology generates signature data
that can provide
image authentication as well as tampering location capability.

\section*{Acknowledgments}

This work was partially supported by the
CNPq (Brazil); and DFAIT and NSERC (Canada).

{\small
\bibliographystyle{IEEEtran}
\bibliography{ffdctwm}
}

\end{document}